\documentclass[aps,twocolumn,superscriptaddress,prb,longbibliography]{revtex4-2}

\usepackage[utf8]{inputenc}
\usepackage[T1]{fontenc}
\usepackage{dcolumn}
\usepackage{bm}
\usepackage{amsmath}
\usepackage{natbib}
\usepackage{graphicx} 
\usepackage{tabularx}
\usepackage{multirow}
\usepackage{rotating} 
\usepackage{makecell}
\usepackage{parskip}
\usepackage{amssymb}
\usepackage{textgreek}
\usepackage[normalem]{ulem}
\usepackage{hyperref}
\hypersetup{colorlinks=true, linkcolor=blue, citecolor=blue, urlcolor=blue}

\usepackage{float}
\usepackage[english]{babel}
\usepackage{tabularx}
\usepackage{booktabs}
\usepackage{lipsum}
\usepackage{graphicx}
\usepackage{siunitx}

\usepackage{lineno}
\newcommand{\bea}{\begin{eqnarray}}
\newcommand{\eea}{\end{eqnarray}}

\def\iMAX{$i$-MAX}

\begin{document}

\title{Magnetic Order and Magneto-Elasticity in the Electronic Excitations of Gd-\iMAX}

\author{Kartik Panda}
\affiliation{Department of Physics, Ariel University, Ariel 4070000, Israel}

\author{Daniel Potashnikov}
\affiliation{Department of Physics, Nuclear Research Centre-Negev, P.O. Box 9001, Beer Sheva 84190, Israel}
\affiliation{Department of Material Science and Engineering, Faculty of Engineering, Tel Aviv University, Tel Aviv 6997801, Israel.}

\author{Asaf Pesach}
\affiliation{Department of Physics, Nuclear Research Centre-Negev, P.O. Box 9001, Beer Sheva 84190, Israel}

\author{Maxime Barbier}
\affiliation{Universit\'e Grenoble Alpes, CNRS, Grenoble INP, LMGP, F-38000 Grenoble, France}
\affiliation{European Synchrotron Radiation Facility (ESRF), 71 Avenue des Martyrs, CS 40220, 38043 Grenoble Cedex 9, France}

\author{Anna Eyal}
\affiliation{Faculty of Physics, Technion – Israeli Institute of Technology, Haifa 32000, Israel}

\author{Thierry Ouisse}
\affiliation{Universit\'e Grenoble Alpes, CNRS, Grenoble INP, LMGP, F-38000 Grenoble, France}

\author{Amit Keren}
\affiliation{Faculty of Physics, Technion – Israeli Institute of Technology, Haifa 32000, Israel}

\author{Nimrod Bachar}
\email{nimib@ariel.ac.il}
\affiliation{Department of Physics, Ariel University, Ariel 4070000, Israel}


\begin{abstract}

We report the investigation of electronic collective modes in rare-earth-based magnets (Mo$_{2/3}$RE$_{1/3}$)$_2$AlC (also known as RE-\iMAX\ phases), where RE=Gd, Yb, and Dy, using single crystal samples. A detailed investigation of the Raman spectra of Gd-\iMAX\ samples at low temperatures, with a focus on the phonon behavior in relation to the antiferromagnetic (AFM) phase transition at 26 K is presented. Significant shifts in the central frequencies of several low-frequency phonon modes were observed below 25 K, correlating with the N\'{e}el transition. Integrated Raman intensity measurements indicated a reduction in the electronic background below the AFM transition temperature, suggesting the opening of a magnetic gap. Our analysis showed no new phonon modes. Therefore, we do not see any indication of a Brillouin zone folding of phonon mode to the $\Gamma$-point in our measurement. However, the hardening of all phonon modes at low temperatures points to a strong spin-phonon coupling effect. Using a temperature-dependent model of phonon frequency, we determined the spin-phonon coupling constant $\lambda$ to be less than 0.1 cm$^{-1}$ for all frequencies, which is of the same order of magnitude as found in other antiferromagnetic materials such as MnF$_{2}$ and FeF$_{2}$~\cite{Lockwood1988} with $T_N=68~K$ and $T_N=78~K$, respectively, but significantly lower than that of $CuO$ with $T_N=213~K$.

\end{abstract}

\maketitle

\section{Introduction}

The exfoliation of two-dimensional (2D) van der Waals (vdW) ferromagnets, such as CrI$_3$ \cite{Huang2017}, and their introduction into multilayer heterostructures contribute considerably to the search for new physical phenomena such as bosonic Dirac excitations \cite{fransson2016magnon,pershoguba2018dirac,chen2018topological,chen2021magnetic,sun2023interacting}, interlayer interaction control via stacking order \cite{klein2019enhancement}, and engineered spin and valleytronics \cite{zhong2017van}. The finding of air-stable vdW 2D magnets \cite{ziebel2024crsbr} is expected to accelerate using 2D ferromagnets in research and technology~\cite{Yang2020}. In contrast, 2D antiferromagnets are less popular, and their possible application is unclear.

\par

The main fundamental question is associated with the Mermin-Wagner theory and its limit on the presence of an order parameter in dimensions lower than 3~\cite{Gibertini2019}. However, due to various spin Hamiltonian models and inherent crystal anisotropies, most materials will preserve the magnetic order parameter in some form.  Up to date, only monolayers of FePS$_3$ maintain the antiferromagnetic state as in the bulk, leaving the N\'{e}el temperature intact because of the discrete spin configuration of an Ising-like Hamiltonian \cite{Wang2016,Lee2016}. 

\par 

Consequently, much effort and interest have recently been drawn into the nano lamellar carbides and nitrides, known as MAX phases~\cite{Petruhins2019,Dahlqvist2022}. In the acronym MAX, M is a metal, A stands for Al or other A-group elements, X is C (Carbides) or N (Nitrides). M can be constructed from two metals, M${^1}$ and M${^2}$. When M${^1}$ and M${^2}$ are ordered in-plane, the compound is called \iMAX\ \cite{tao2017two}. 

\par 

Ferromagnetism and antiferromagnetism are properties associated with a collective spin behavior through, e.g., the exchange interactions between magnetic moments. Therefore, expecting collective magnetic modes in the dynamic conductivity spectrum is natural. Magnons are well-known collective modes that can be observed in the energy range of about few $\mu$eV to few meV, as in the case of canonical transition metal fluorides and oxides~\cite{Ohlmann1961,Talbayev2004}. The magnons identified recently in the absorption spectrum of FePS$_3$ in the THz range are in excellent agreement with inelastic neutron scattering measurements~\cite{Wildes2012}. The magnetic phase is associated with an order parameter and an energy gap; therefore, it changes the electronic background around the various collective modes, charge (phonons), and spin. Moreover, hybrid spin-charge collective modes are also expected due to these materials' strong coupling between charge and spin degrees of freedom. 

Optical spectroscopy is one of the significant tools that can access and directly probe the collective behavior in condensed matter physics~\cite{Basov2011}. Electronic interactions that lead to exotic quantum phases of matter, such as superconductivity, density waves, and Mott insulating states, have been studied extensively and successfully using optical spectroscopy in various materials and a wide frequency range~\cite{Basov2005,Basov2011}. Since optical probes such as infrared and Raman spectroscopy can be utilized at the microscopic level, it is also helpful to use them as a direct probe of collective excitations in 2D magnets~\cite{Burch2018}. For example, the optical Kerr effect and Raman microscopy were used extensively to identify the magnetic state via the spontaneous symmetry breaking and appearance of magnetic collective modes in monolayers of 2D magnet compounds.

This work uses Raman spectroscopy to probe the electronic collective modes of \iMAX\ single crystals with the chemical formula (Mo$_{2/3}$RE$_{1/3}$)$_2$AlC, where RE=Gd, Yb, and Dy. We compare the collective excitations' spectrum of the different phases with the expected values obtained by DFT calculations. We study the Raman spectrum at low temperatures, above and below the magnetic phase transition in Gd-\iMAX\ samples. We observe a clear signature of the phase transition close to T$_N$ (as measured by magnetometry probes) in the Raman shift spectrum. The electronic background is removed while the central frequency of the vibrational modes is shifting. We ascribe this behavior to forming an AFM gap and changing the modes' binding forces due to introducing the superexchange energy between neighboring magnetic moments. 

\section{Experimental details}

Gd-\iMAX\ single crystals were synthesized by using a growth flux method at high temperature, similar to that described in~\cite{Barbier2022}. Gd, Mo, and Al pellets were mixed with atomic fractions $x_{\text{Gd}}=0.644$, $x_{\text{Al}}=0.276$, and $x_{\text{Mo}}=0.08$ inside an induction-heated graphite crucible. Carbon was provided by partially dissolving the crucible walls at high $T$, and the crucible was sealed in order to avoid Gd and Al evaporation. After reaching a maximum temperature $T=1800\,^\circ\text{C}$ in an $Ar$ atomsphere and ambient pressure, the crucible was slowly cooled down to $T=1000\,^\circ\text{C}$ during 7 days, and then left to cool down freely. Crystals were extracted from the flux by oxidizing the latter at room temperature in a chamber equipped with air-bubbling. Crystals were flakes with typical lateral size of a few $100\,\mu\text{m}$.

\par 

The temperature-dependent DC susceptibilities, $\chi$(T), were obtained from Zero-field-cooled (ZFC) magnetization measurements conducted at the Quantum Material Research Center in the Technion using the Quantum Design MPMS3 system. The measurements were carried out at a low magnetic field ($\mu_0H$) of 1 kOe, aligned along the \textit{c*} axis of RE-\iMAX\ (RE = Gd, Dy, Yb) single crystals. Gd-\iMAX\ and Dy-\iMAX\ samples were also measured at a high magnetic field, $\mu_0H$ = 70 kOe.
\par 

Raman spectroscopy measurements were performed using a Labram HR Evolution Horiba system combined with a CryoVac liquid helium flow cryostat. A 532 nm laser was used as the excitation source and was focused on the sample (inside the cryostat) through a 63× Olympus objective. The scattering light from the sample was collected by the same objective, passed through the analyzer, sent to a Czerni–Turner spectrometer equipped with either a 600 or 1800 grooves per mm grating, and was detected by a liquid nitrogen-cooled CCD-array.

\par 

The Raman spectra were obtained with an 1800 grating in between room temperature and 4~K for all RE = Gd, Dy, Yb samples to study the detailed vibrational modes spectra compared with previous DFT calculations~\cite{Champagne2019}. Further, we have studied the temperature dependence of the Raman spectra of a Gd-\iMAX\ sample in a temperature interval from 4~K to 50~K in steps of 1~K. A custom-made alignment setup, including a motorized XYZ stage and a machine learning-based image analysis algorithm, was used to align the visible sample image, both planar and focal, each time temperature was stabilized and before a scattered light spectrum was collected. This allowed a careful and absolute comparison of the Raman spectra, particularly the electronic background's contribution as a function of temperature.

\par 

The Raman spectrum was fitted using RefFit software and a specialized Raman model (model 113)~\cite{Kuzmenko2005}. The high-resolution data (i.e., obtained by the 1800 groves per mm grating) was fitted in two approaches: The first approach included the full spectrum from -500 cm$^{-1}$ to 350cm$^{-1}$, i.e., including the anti-stokes excitations. The second approach focused only on the stokes excitation from 45cm$^{-1}$ to 350cm$^{-1}$. Both methods had the same vibrational modes and electronic background temperature dependence.    

\section{Results}

\begin{figure}
    \centering
    \includegraphics[width=0.9\linewidth]{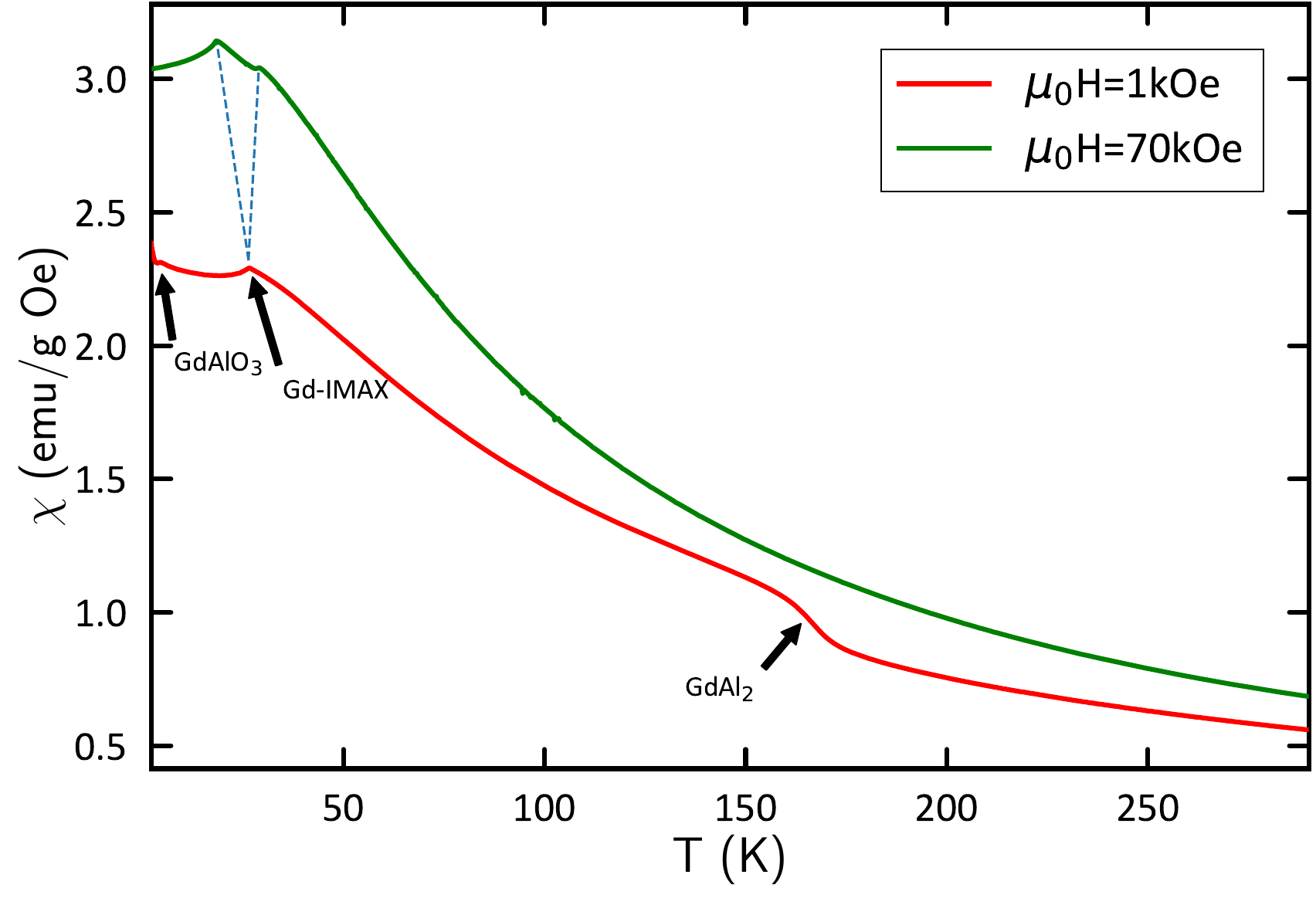}
    \caption{Temperature dependent DC magnetization of single crystal Gd-\iMAX\ sample at 0.01~T (red) and 7~T (green). The applied field is aligned along \textit{c*} axis of the crystal.}
    \label{fig:Gd_MvsT}
\end{figure}

The temperature-dependent susceptibility $\chi$(T) of Gd-\iMAX\ single crystal sample is shown in Fig.~\ref{fig:Gd_MvsT} at 1~kOe and 70~kOe. At 1~kOe, two maxima of $\chi$(T) are observed at 4.57~K, and 26.6~K, indicating the N\'{e}el temperatures of GdAlO$_3$~\cite{Cashion1968} and Gd-\iMAX \cite{Tao2019}, respectively, below which an antiferromagnetic (AFM) order exists. These features of $\chi$(T) are followed by a change of slope at 167~K, which is related to the Curie temperature of the paramagnetic (PM) -ferromagnetic (FM) phase transition in GdAl$_{2}$~\cite{Levin2001}. At 70~kOe, however, the contribution of the magnetic impurities, GdAlO$_{3}$ and GdAl$_{2}$, to $\chi$(T) is dissipated while the AFM transition in Gd-\iMAX\ is two-folded, suggesting two magnetic phase transitions, at 18.5~K and 29~K.

\begin{figure}
    \centering
    \includegraphics[width=0.9\linewidth]{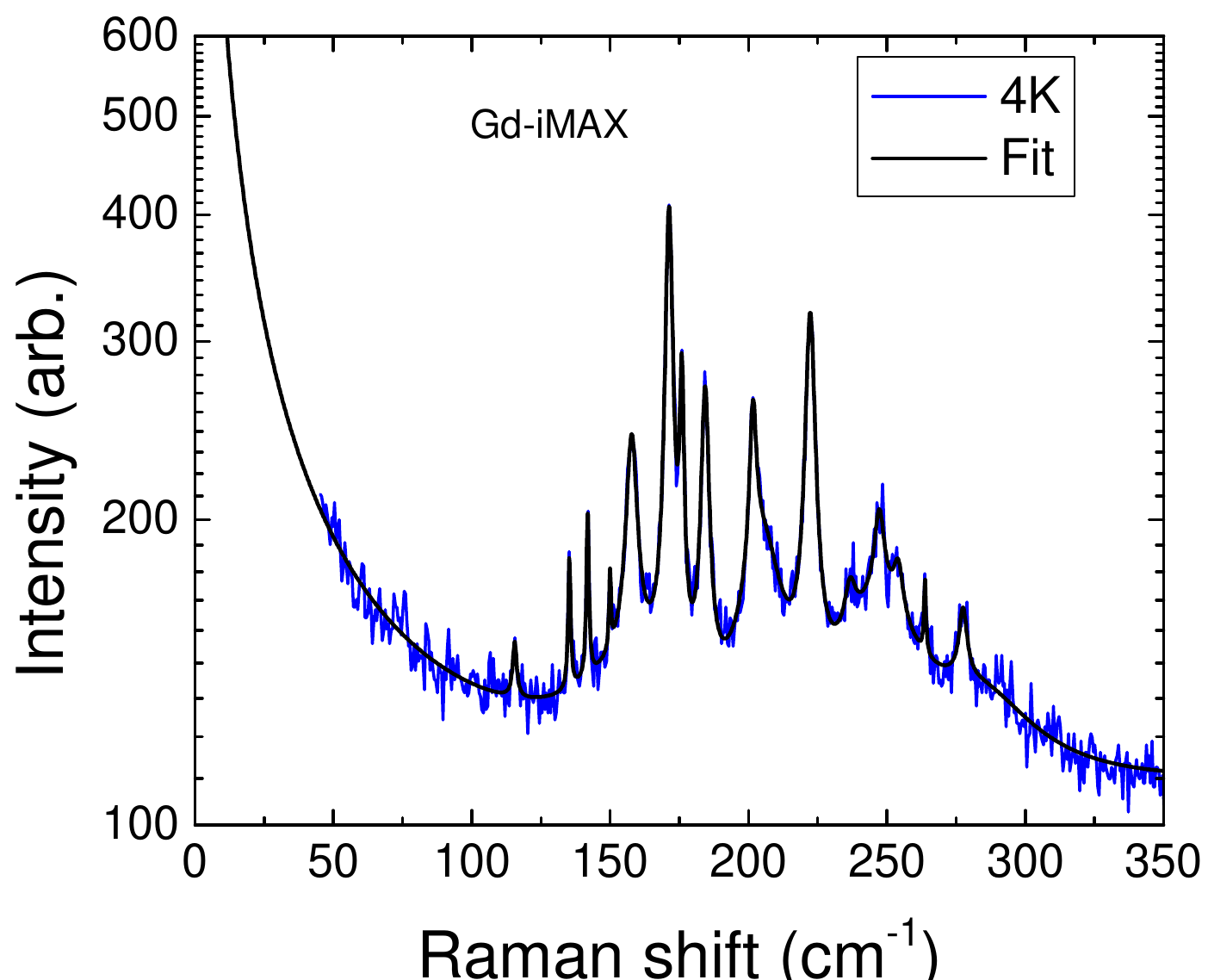}
    \caption{Raman spectra of Gd-\iMAX sample at 4 K. The solid line shows the fitting to the data.}
    \label{fig:GdData4K}
\end{figure}

\begin{figure}
    \centering
    \includegraphics[width=\linewidth]{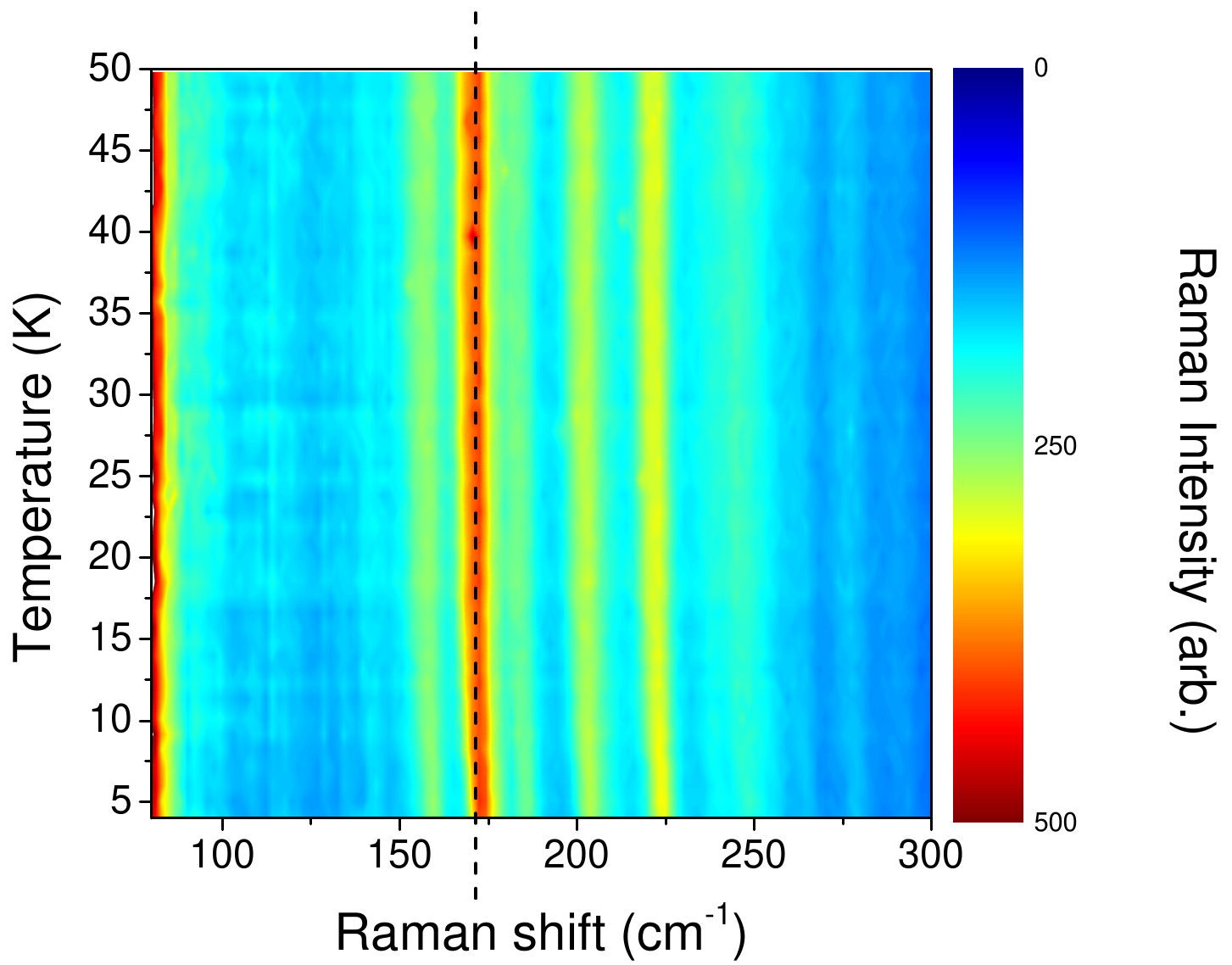}
    \caption{Contour plots of Raman spectra as a function of temperature. The dotted line is the guide to the eye. It shows clearly the shift phonon frequency with temperature.}
    \label{fig:Raman2D}
\end{figure}

\begin{table}
    \centering
     \caption{Phonon modes in Gd-, Dy-, and Yb-\iMAX\ samples. The experimental values were obtained from the fit to a Raman scattering model in RefFit. The theoretical values for Rare Earth (RE) = Gd and Dy, along with the phonon symmetries, were adopted from Ref.~\cite{Champagne2019}.}
    \begin{tabular}{cc|cc|cc|c}
        \hline\hline 
         &  & \multicolumn{2}{c|}{Gd} & \multicolumn{2}{|c|}{Dy} & \multicolumn{1}{|c}{Yb} \\
       Symmetry   & Atoms      & Exp.         & Th.   & Exp.  & Th.   & Exp.   \\
         \hline\hline
       A$_g$(1)   & RE+Mo+Al   & --           & 111.2 & --    & 110.1 &        \\
       B$_g$(1)   & RE+Mo+Al   & 115.5        & 112   & 114.0 & 110.8 & 111.3  \\
       B$_g$(2)   & RE         & --           & 123.2 & 123.3 & 120.7 & --     \\
       A$_g$(2)   & RE         & 135.3        & 130.3 & 132.2 & 127.3 & 131.1 \\
       B$_g$(3)   & RE         & 141.9        & 137.7 & 138.4 & 134.3 & 133.8  \\
       A$_g$(3)   & RE+Mo      & 150.0        & 145.9 & 143.8 & 139.8 & --     \\
       A$_g$(4)   & RE+Mo(+Al) & 157.8        & 157.9 & 157.3 & 156.4 & 154.6  \\
       B$_g$(4)   & RE+Mo+Al   & --           & 165.5 & --    & 162.0 & --     \\
       A$_g$(5)   & RE+Mo+Al   & --           & 169.4 & --    & 168.1 & --     \\
       B$_g$(5)   & Mo         & 171.3        & 169.0 & 174.0 & 171.8 & 169.8  \\
       A$_g$(6)   & Mo         & 175.9        & 177.8 & --    & 179.6 & 177.8  \\
       B$_g$(6)   & Mo         & 184.3        & 185.3 & 185.8 & 185.6 & 188.5  \\
       A$_g$(7)   & Al(+Mo)    & --           & 194.7 & --    & 186.1 & --     \\
       B$_g$(7)   & Mo         & 201.6        & 195.2 & 204.4 & 197.1 & 206.2  \\
       A$_g$(8)   & Mo         & --           & 197.2 & --    & 198.6 & --     \\
       B$_g$(8)   & Mo+Al      & 205.1        & 204.7 & --    & 206.4 & 212.6  \\
       A$_g$(9)   & Mo+(Al)    & 222.3        & 218.5 & 224.4 & 220.0 & 226.6  \\
       B$_g$(9)   & Mo+Al      & 236.7        & 233.2 & 240.0 & 235.4 & 243.4  \\
       A$_g$(10)  & Mo+Al      & --           & 244.0 & --    & 246.0 & --     \\
       B$_g$(10)  & Mo+Al      & 247.3        & 245.2 & 249.8 & 247.3 & 253.7  \\
       A$_g$(11)  & Mo+Al      & --           & 250.2 & --    & 251.2 & --     \\
       B$_g$(11)  & Mo+Al      & 254.1        & 254.1 & --    & 260.2 & --     \\
       B$_g$(12)  & Mo+Al      & 263.7        & 266.2 & 269.6 & 271.5 & --     \\
       A$_g$(12)  & Mo+Al      & --           & 267.7 & --    & 273.8 & --     \\
       A$_g$(13)  & Mo+Al      & 277.5        & 270.5 & 282.8 & 275.9 & 291.6  \\
       B$_g$(13)  & Mo+Al      & --           & 282.5 & --    & 290.1 & 310.9  \\
       B$_g$(14)  & Al         & --           & 353.1 & --    & 361.9 &        \\
         \hline
    \end{tabular}
    \label{tab:phonons}
\end{table}

The Raman spectrum of a Gd-\iMAX\ sample at 4~K is shown in Fig.~\ref{fig:GdData4K}. The data was fitted to capture the various vibrational modes, as detailed in Table~\ref{tab:phonons}. The temperature dependence of the Raman intensity (raw data) is shown in Fig.~\ref{fig:Raman2D}. Several phonons in the measured range, particularly low-frequency modes of 130 to 250 cm$^{-1}$, are changing their central frequency at low temperatures and below 25~K. Fig.~\ref{fig:Raman2D} shows a guide to the eye (dashed line) of the central frequency of the strongest phonon in this range, namely the A$_g$(6) mode at 171~cm$^{-1}$, with a clear deviation of $\omega_0(50K)$ can be seen at low temperatures. We note that the temperature dependence spectra were taken with a lower-density grating to increase the signal intensity at the expense of frequency resolution of the A$_g$(6), which is seen as two peaks in the spectrum measured with the high-density grating (Fig.~\ref{fig:GdData4K}).

\par 

\begin{figure}
    \centering
    \includegraphics[width=0.75\linewidth]{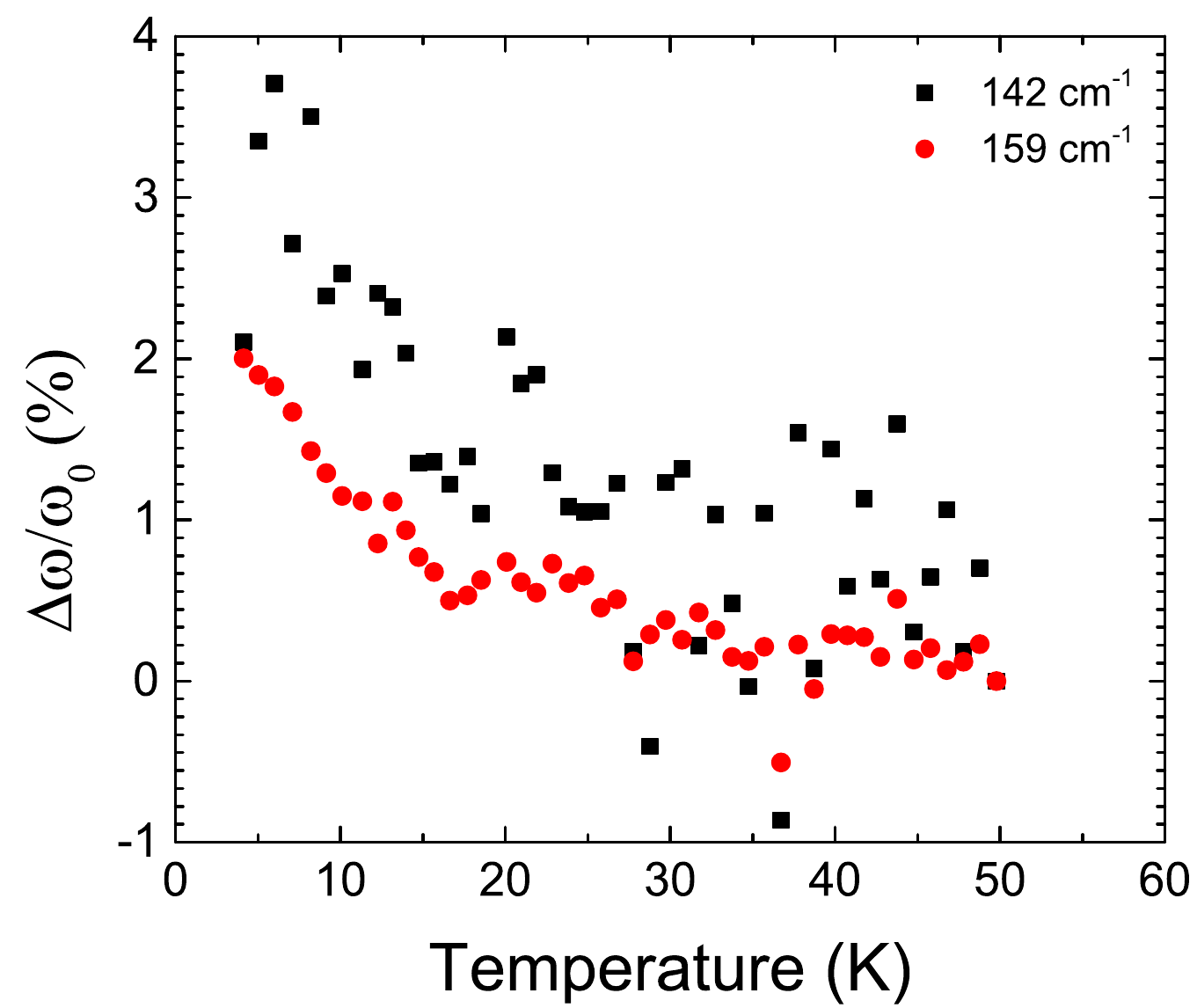}
    \includegraphics[width=0.75\linewidth]{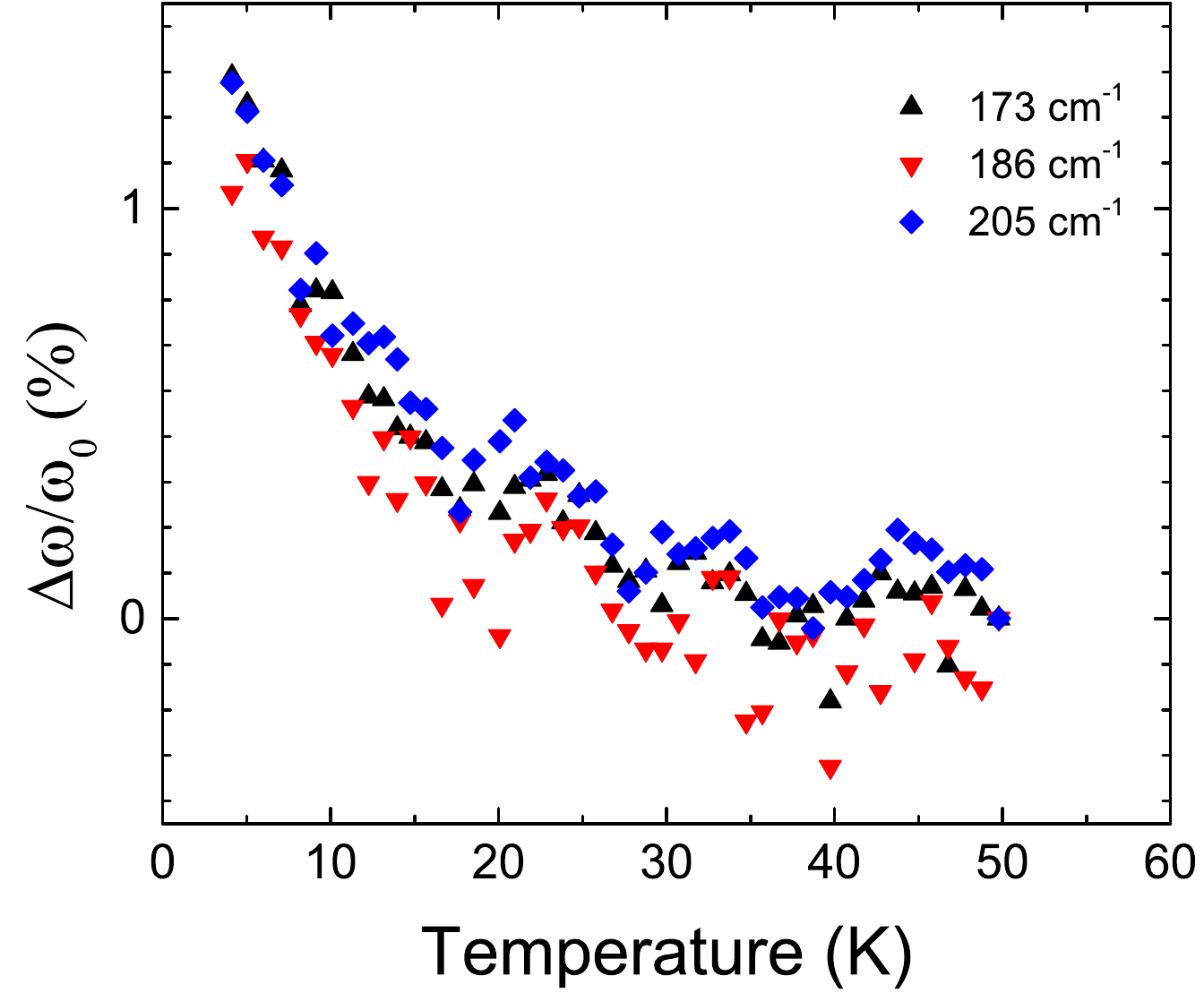}
    \includegraphics[width=0.75\linewidth]{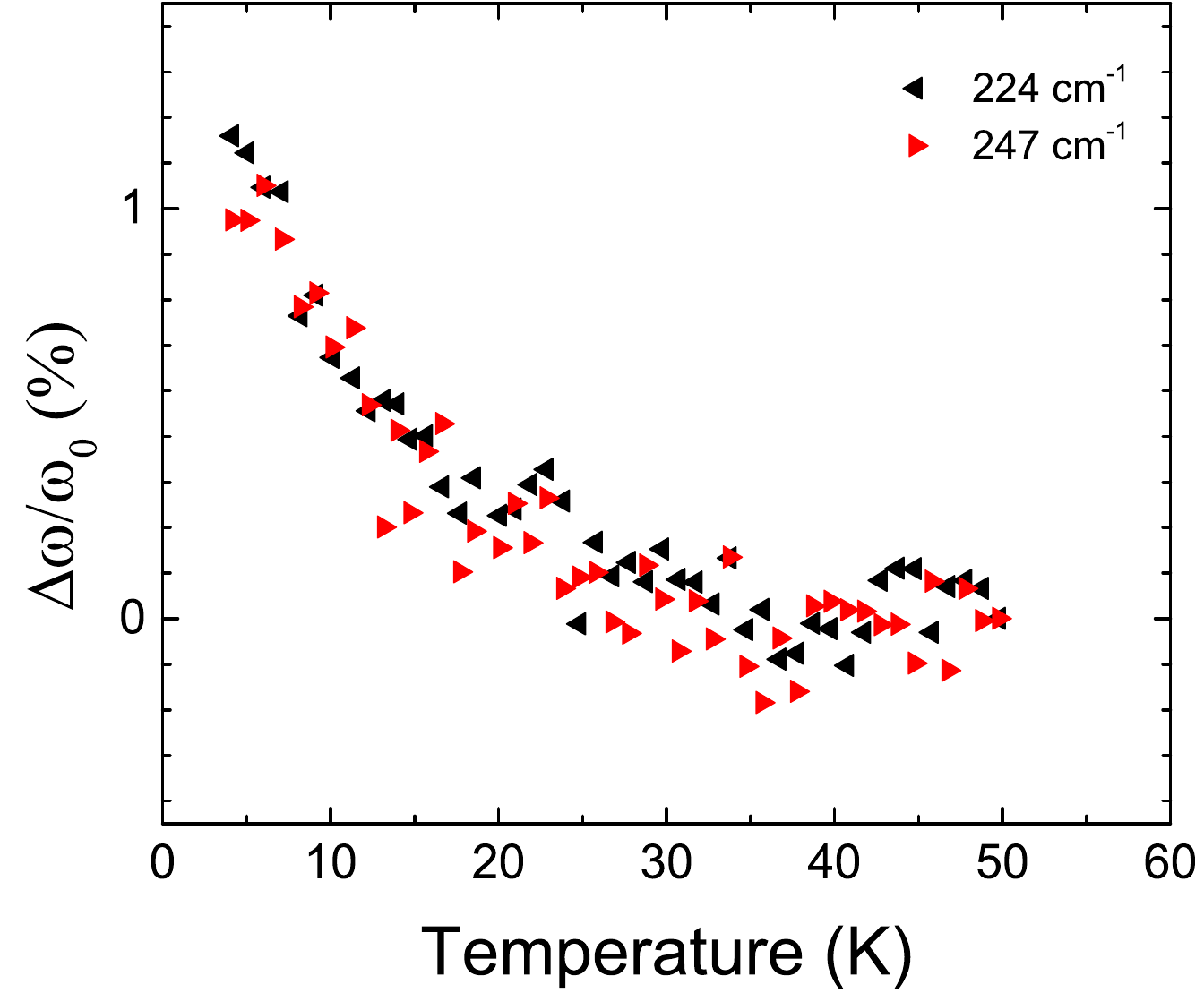}
    \caption{The relative change of the frequency normalized to the central frequency at base temperature as a function of temperature where $\Delta \omega = \omega-\omega_{0}$ (a) for 142 cm$^{-1}$ and 159 cm$^{-1}$ (b) for 173 cm$^{-1}$, 186 cm$^{-1}$ and 205 cm$^{-1}$ (c) 224 cm$^{-1}$ and 247 cm$^{-1}$. }
    \label{fig:Gd_Omega0T}
\end{figure}

Fig.~\ref{fig:Gd_Omega0T} shows the temperature dependence of the central frequency of a selection of phonons in the measured spectrum of the Gd-\iMAX\ sample as given by the fit analysis. $\omega_0(T)$ shows a deflection point from the high-temperature static frequency in the range of 20 to 30~K and in close agreement with the N\'{e}el transition of Gd-\iMAX\ at 26~K. In all cases, $\omega_0$ is shown to increase as the temperature is lowered, while the increase is smaller for high-frequency vibrational modes (about 0.5\% compared to more than 2\% for the low-frequency modes). 

To study the full effect of the AFM transition, we used the spectral weight method of optical conductivity in our measurements. Although this method here does not give any quantitative property, such as the number of electrons per site, as in the case of the $f$ sum rule, it is still possible to understand how the electronic background changes through the phase transition. Since our Raman data is obtained precisely with the same measurement parameters, it is possible to compare the Raman intensities as a function of temperature quantitatively. 

\par

\begin{figure}
    \centering
    \includegraphics[width=0.9\linewidth]{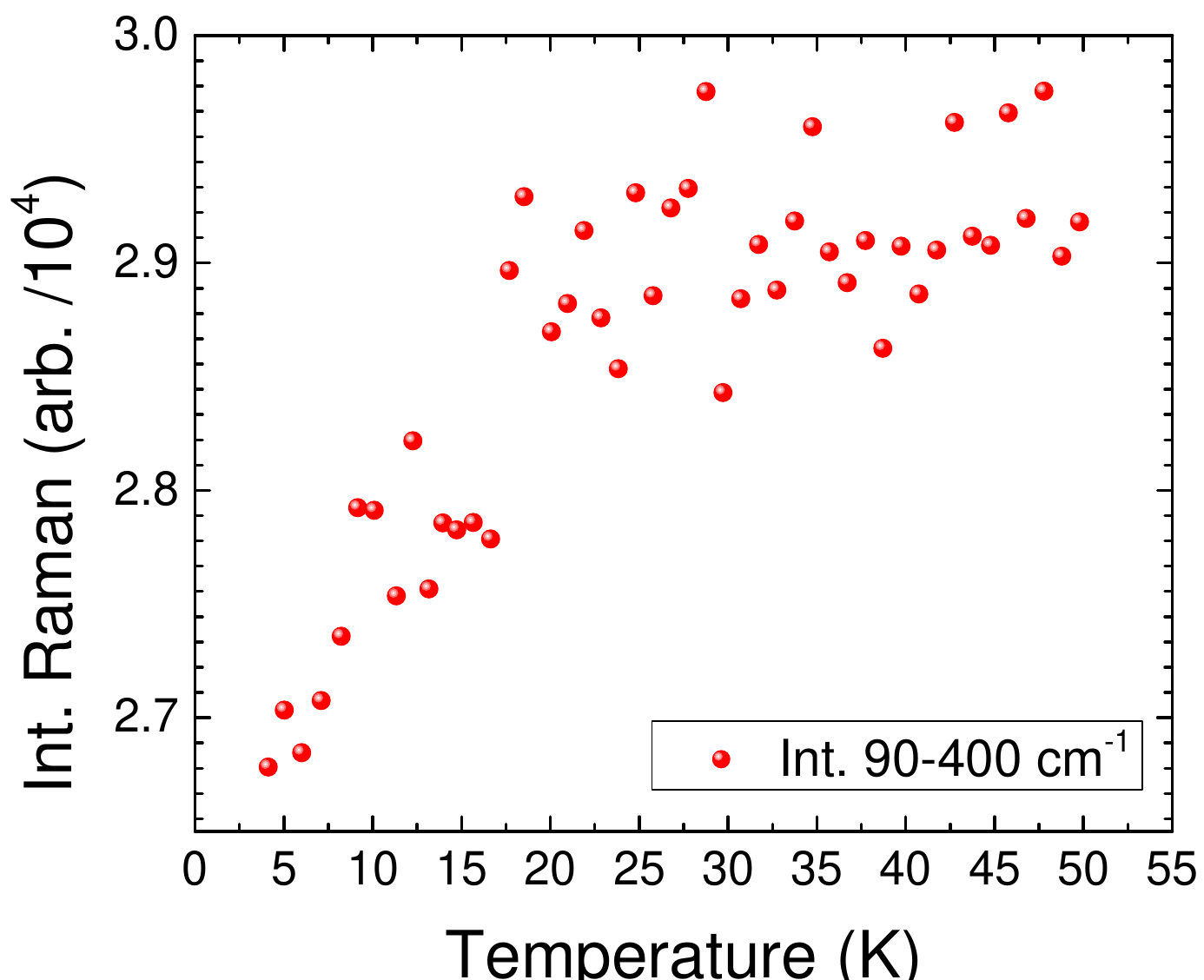}
    \caption{The integrated Raman spectrum between 90~cm$^{-1}$ and 400 $cm^{-1}$ as a function of temperature. }
    \label{fig:RamanInt}
\end{figure}

Fig.~\ref{fig:RamanInt} shows the result of the integrated Raman spectrum between 90~cm$^{-1}$ and 400 $cm^{-1}$. The total Raman intensity decreases below 25~K, suggesting that the electronic background is lower below the AFM phase transition.  

\section{Discussion}

Raman spectrum is typically used to detect lattice vibrational modes and other collective modes, such as (bi-)magnons. In particular, Raman spectroscopy is coupled to the even collective modes superimposed on the electronic background. Phase transitions to a state with an order parameter, such as superconductivity, (anti)ferromagnetism, or spin density waves, are accompanied by the opening of a spectroscopic gap typically seen as a reduction of the dynamic conductivity. Therefore, such effects should also affect the electronic background measured by the Raman spectroscopy apparatus. Indeed, the "spectral weight"-like analysis shown in Fig.~\ref{fig:RamanInt} shows that the electronic background is reduced below T$_N$ of 26~K for the Gd-\iMAX\ sample. Hence, our Raman scattering data shows a spectroscopic signature of the magnetic gap and the magnetic phase transition.   

\par

In some cases, along with the effects involving the opening of a gap, and particularly those involving a magnetic phase transition, there is a doubling of the unit cell in the magnetic state. As a result, a Brillouin Zone folding of high-momentum optical phonon branches into the zero momentum $\Gamma$ point is responsible for the appearance of new modes in the spectra~\cite{Hashemi2017,Wang2016,Lee2016}. As shown in Fig.~\ref{fig:GdData4K}, no new modes appear in our data down to the noise level of our measurements. 

\par 

In general, lattice vibrations are mainly characterized by charge effects, while magnetic forces are negligible. However, magnetic materials could have magneto-elastic effects due to the coupling between the spin and lattice degrees of freedom. This results in hybrid spin-charge phonon modes as in other 2D magnetic materials, particularly van der Waals magnets~\cite{Wyzula2022}. The manifestation of the coupling can be detected by a change in the phonon's central frequency along with Kerr/Faraday angle rotation around the phonon's frequency~\cite{Gong2017,Huang2017}. In extreme cases of high magnetic fields, there are also effects of avoided crossing between the modes, such as a magnon and a phonon, which are close to each other in frequency. Indeed, our data suggests that such coupling also exists in the Gd-\iMAX\ sample due to the low-temperature evolution of the phonons' central frequency below $T_N$ of 26~K. 

\par 

Generally, a shift in the phonon frequency is expected for magnetoelasticity or when there is a coupling to the electronic background. Naively speaking, in an antiferromagnet, one wishes to have a weaker force constant between atoms in the cold lattice model having opposite spins compared to atoms having parallel spins as in a ferromagnet. Moreover, suppose the electronic background counter oscillates in resonance with the lattice modes, such as a Fano-shaped phonon. In that case, the reduction of the background should reduce the force constant of the phonon itself. In both cases, one should expect the phonon's central frequency to shift when entering the antiferromagnetic state, where the spin-phonon coupling constant will determine whether the phonon hardens or softens as a function of decreasing temperature.  

\begin{figure}
    \centering
    \includegraphics[width=0.85\linewidth]{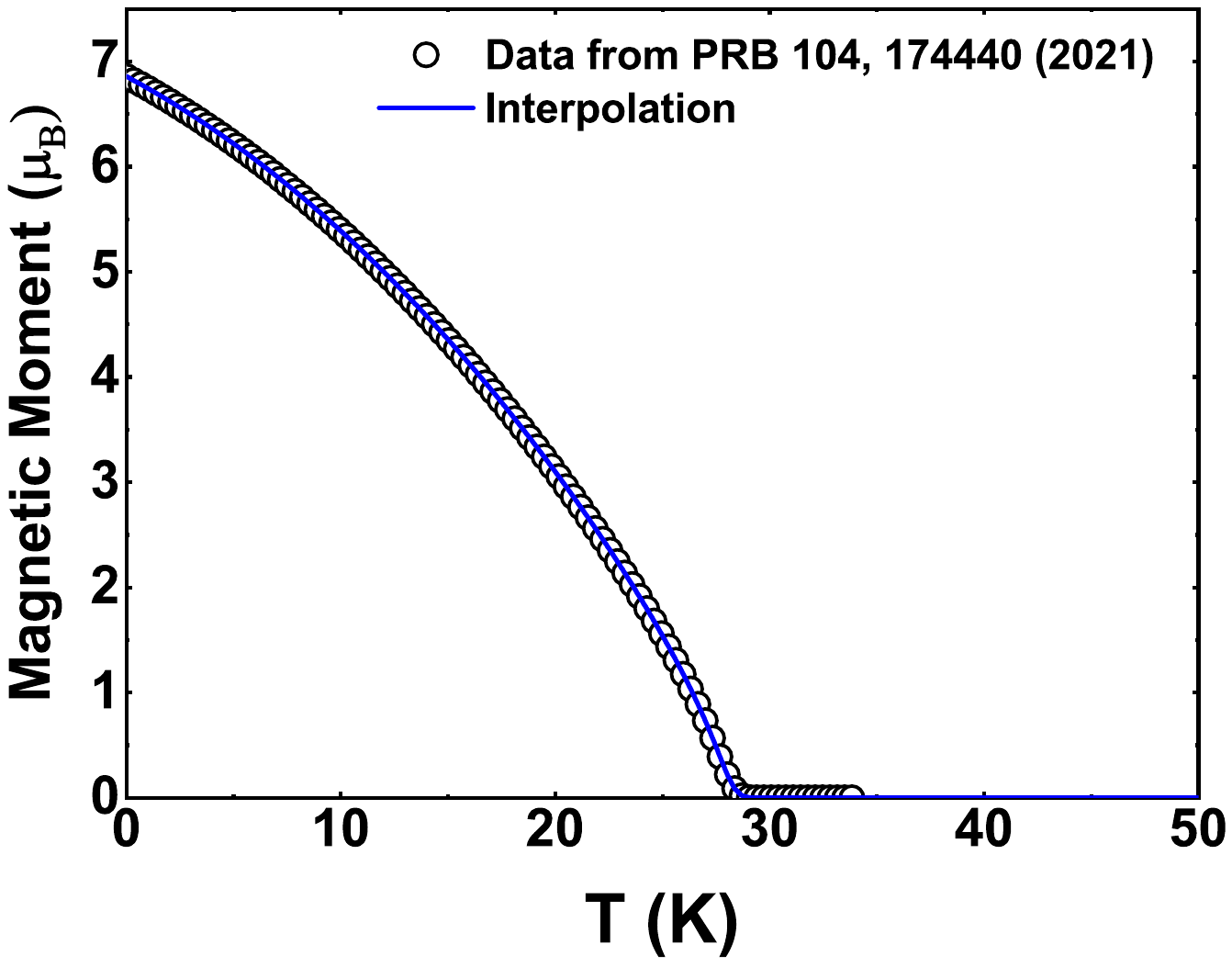}
    \includegraphics[width=0.9\linewidth]{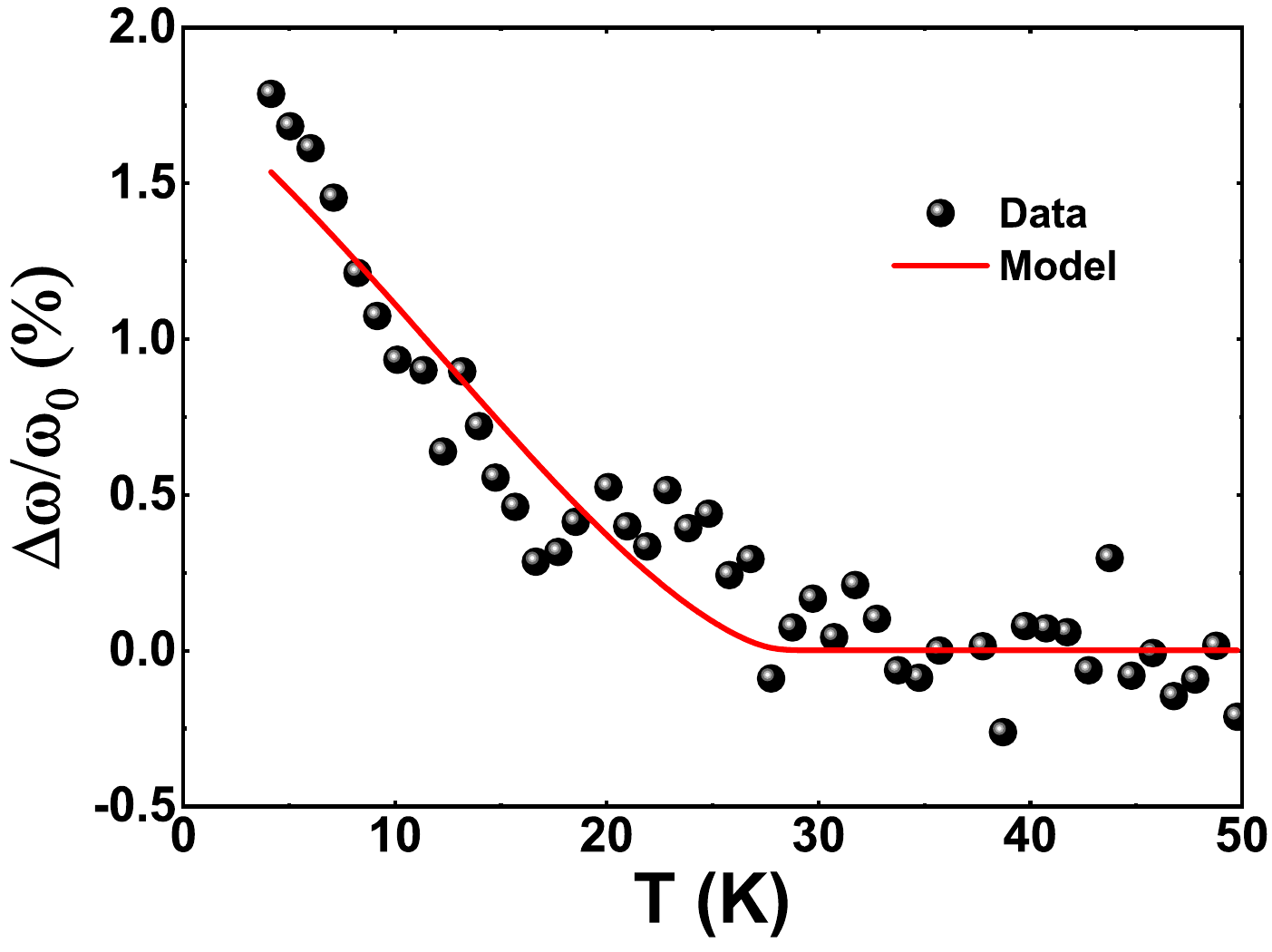}
    \caption{Top panel: Magnetic moment as a function of temperature. Data is taken from Ref.~\cite{Potashnikov2021}. Data is extrapolated up to 50 K. Lower panel: The relative change of the frequency normalized to the central frequency (156 cm$^{-1}$ at base temperature as a function of temperature where $\Delta \omega = \omega-\omega_{0}$. The solid red line shows the fitting using Equation~\ref{eq:fit}.}
    \label{fig:Omega_fit}
\end{figure}

\begin{figure}
    \centering
    \includegraphics[width=0.9\linewidth]{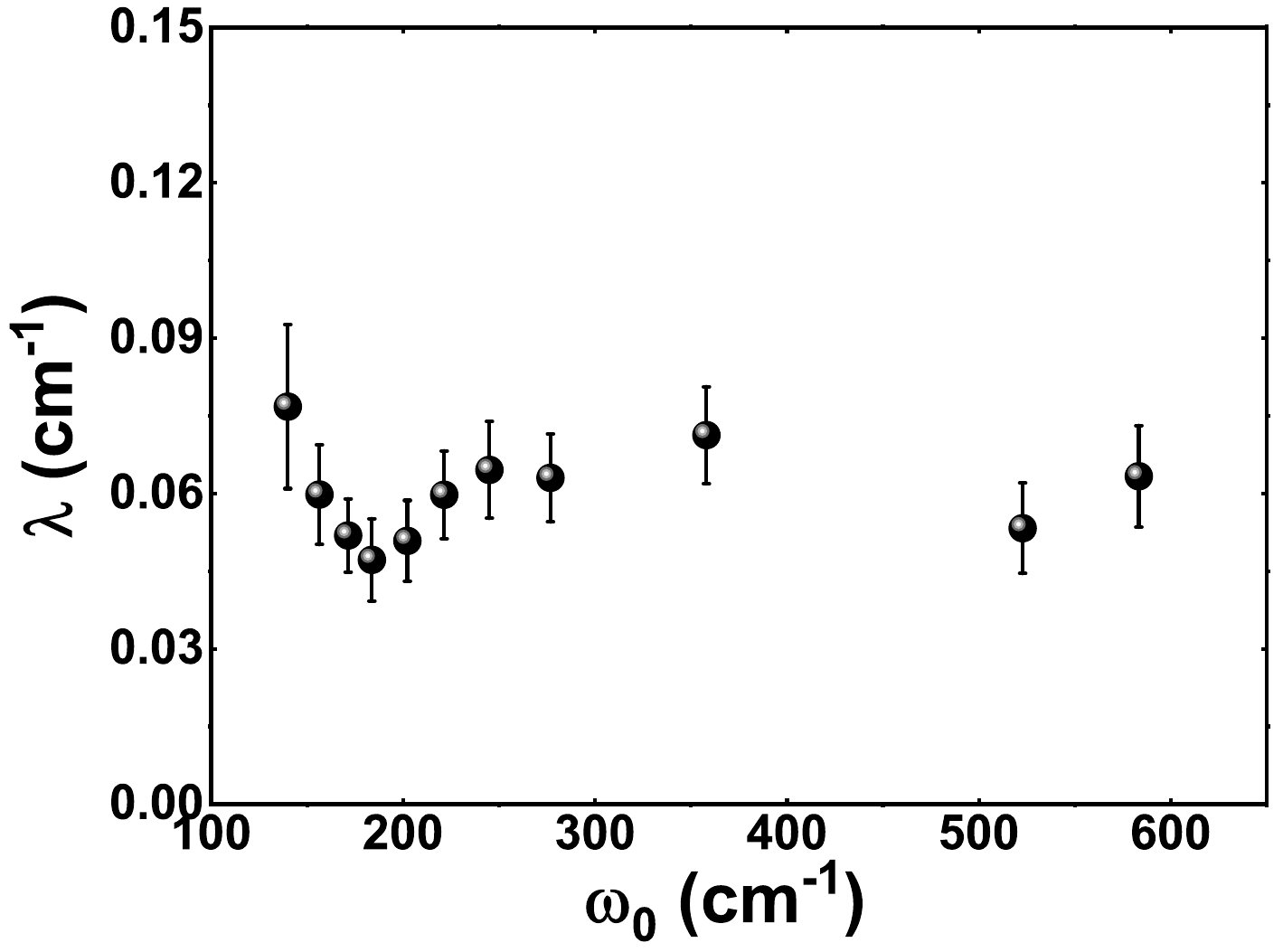}
    \caption{Spin-phonon coupling coefficients (in cm$^{-1})$ for different central frequency}
    \label{fig:Lambda}
\end{figure}

\par 

The frequency shift in the phonon frequency can be described by the Hamiltonian of the atomic lattice energy term, $H_l$ and the spin-lattice energy term:

\begin{equation}
    H = H_l + \sum\limits_{i,j} J_{ij} S_i S_j
\end{equation}

where $J_{ij}$ is the exchange interaction constant between the spins at sites $i$ and $j$ and $S$ is the spin operator. $J$ can be expanded up to the second order in the displacement of the nearest neighboring spins $C_{m}S^2$ and introduced back to the Hamiltonian. As a result, the phonons central frequency will shift due to the magneto elastic coupling. Here $C_m$ is proportional to the spin-phonon coupling constant and $S^2=\langle S_{i} S_{j}\rangle$ $\langle S_{i} S_{j}\rangle$ is the statistical average value of the neighboring spins~\cite{Lockwood1988,Chen1995}. The central phonon frequency $\omega(T)$ can be then expressed as:

\begin{equation}
    \omega(T) = \sqrt{\omega_{0}^2+C_{m}M(T)^2}
    \label{eq:fit}
\end{equation}

where $\omega_{0}$ is the phonon frequency in the absence of spin-phonon coupling, observed above $T_{N}$, and $M(T)$ denotes the magnetic moment as a function of temperature for the Gd-\iMAX sample. Due to the magneto elastic coupling, $C_m$ can take a positive (negative) value depending on the exact hardening (softening) of the phonon's spring constant in the harmonic approximation. 

\par

We extrapolated $M(T)$ from Ref.~\citep{Potashnikov2021} to this work's corresponding temperature measurement range as shown in Fig.~\ref{fig:Omega_fit}. Eqn.~\ref{eq:fit} was then used to fit the temperature dependence of the central frequency where the values of $\omega_{0}$ and $C_{m}$ were taken as the fit parameters. Given that the sample temperature is slightly elevated due to the Laser power, we evaluated the goodness of fit, ($\chi^2$), for a temperature shift, $\Delta T$, ranging from -5~K to +5~K. The minimum of $\chi^2$ was found within the range -2 K < $\Delta T$ < 2 K for all phonons, suggesting a slight temperature shift due to the laser heating. The $\Delta T$ we obtain from the fit is slightly lower than the apparent shift in the transition temperature of the electronic background in the Raman spectrum, which was seen at about 5~K below the actual $T_N$ of the Gd-\iMAX sample. 

\par 

Figure~\ref{fig:Omega_fit} shows an example of the relative change in frequency, $\Delta \omega = \omega - \omega_{0}$, normalized to the central frequency $\omega_{0}$=156~ cm$^{-1}$, as a function of temperature. We also plot in Fig.~\ref{fig:Omega_fit} the fitted model according to Eq.~\ref{eq:fit}. The same fit procedure was done to all phonons showing a frequency shift in their central frequency below $T_N$ to obtain the values of $\omega_{0}$ and $C_{m}$. The shift in the central frequency can be expanded to $\omega(T)=\omega_0+\lambda S^2$, where $\lambda = -C_{m}/2\omega_{0}$ is the spin-phonon coupling coefficient~\cite{Lockwood1988}. Figure~\ref{fig:Lambda} shows the values of the spin-phonon coupling constant as a function of the phonons' central frequency $\omega_{0}$. The value of $\left | \lambda \right |$ was found to be less than 0.1 cm$^{-1}$ for Gd-\iMAX with $T_N\approx27K$, which is of the same order of magnitude as found in other antiferromagnetic materials such as MnF$_{2}$ and FeF$_{2}$~\cite{Lockwood1988} with $T_N=68~K$ and $T_N=78~K$, respectively, but significantly lower than that of $CuO$ with $T_N=213~K$~\cite{Chen1995}.

\section{Conclusions}

In this study, we have investigated the Raman spectra of \iMAX\ samples at low temperatures. We measured the vibrational modes of Gd-, Dy-, and Yb-\iMAX\ single crystals at low temperature and below the antiferromagnetic transition temperature, $T_N$ (in the case of Gd and Dy). Only in the case of Gd-\iMAX\ and since its T$_N$ is substantially higher than the cryostat's base temperature, we have conducted a careful temperature-dependent measurement to obtain the frequency shift of the vibrational modes due to the magnetic phase transition. The phonon modes' central frequency in Gd and Dy \iMAX\ is consistent with previous DFT calculations. Furthermore, our results show no emergence of new modes down to the noise level of our measurement below $T_N$. Therefore, we do not see any indication of a Brillouin zone folding of phonon mode to the $\Gamma$-point in our measurement due to a doubling of the magnetic unit cell if it happens in Gd-\iMAX. The temperature-dependent Raman scattering spectra revealed that several phonon modes exhibit hardening in their central frequencies as the temperature is lowered below $T_N$. The hardening of the phonon modes at low temperatures suggests a magnetoelastic response due to a spin-phonon coupling between the magnetic moments in the Gd-\iMAX\ crystal structure. The spin-phonon coupling constant $\lambda$ is less than 0.1 cm$^{-1}$ for all frequencies, comparable with that of archetype antiferromagnets such as the transition metal fluorides. Concomitantly, the Raman electronic background decreases below the AFM transition temperature. This reduction indicates the order parameter gap opening associated with the magnetic phase transition. 

\par 

In summary, our Raman spectroscopy study highlights the interplay between spin and lattice degrees of freedom in the magnetic order of \iMAX\ single crystals due to the magnetoelastic effects. Therefore, further examination of the collective modes spectrum of other \iMAX\ single crystals, both in Raman and Infrared, as a function of temperature and magnetic field, will be a promising test bed for the spin-phonon coupling mechanism in these materials.

\section*{Acknowledgement}

This research was supported by the Israel Science Foundation (ISF) through project number 666/23 and by the Pazy Research Foundation. N.B. would like to thank J. Teyssier and V. Multian for their help with the Raman spectroscopy setup. N.B. would like to acknowledge useful discussions with P. Barone, C. Faugeras, J. Lorenzana, and D. van der Marel.

\appendix

\section{Magnetic Suscpetibility}

\begin{figure}
    \centering
    \includegraphics[width=0.9\linewidth]{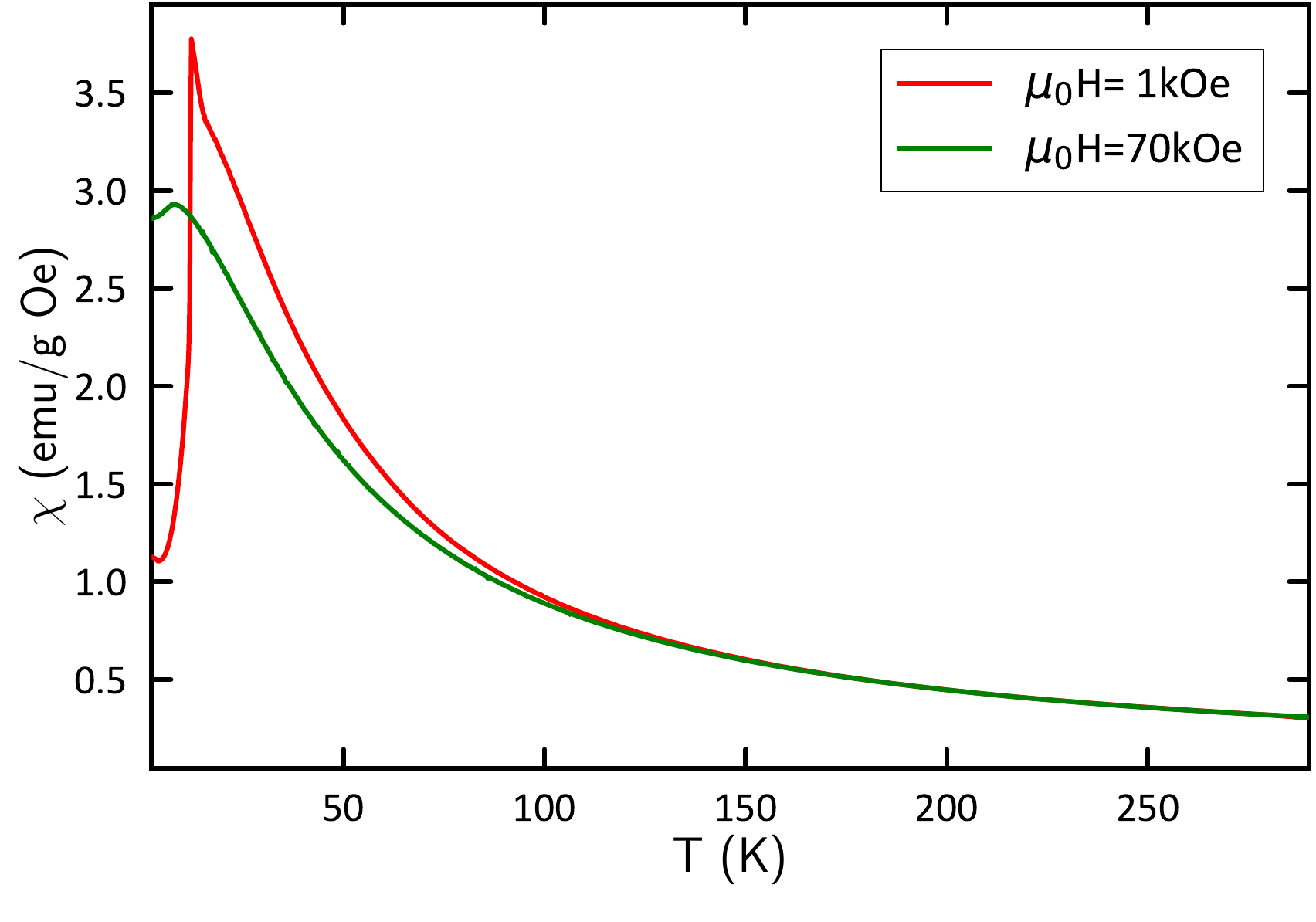}
    \caption{Temperature dependent DC magnetization of single crystal Dy-\iMAX\ sample at 0.01~T (red) and 7~T (green). The applied field is aligned along the \textit{c*} axis of the crystal.}
    \label{fig:Dy_MvsT}
\end{figure}

The temperature dependent susceptibility $\chi$(T) of Dy-\iMAX\ single crystal is shown in Fig.~\ref{fig:Dy_MvsT}, at 1~kOe and 70~kOe. The maximum of $\chi$(T) at 15~K, followed by a sharp drop at 12~K, were already related to a double magnetic phase transition in Dy-\iMAX\, when the external magnetic field is aligned along the $a$ crystal axis~\cite{Barbier2022}. At 70~kOe, a maximum of $\chi$(T) is observed at 8.2~K, suggesting a single PM-AFM transition occurring at the temperature range between 1.9~K and 300~K. 

\begin{figure}
    \centering
    \includegraphics[width=0.9\linewidth]{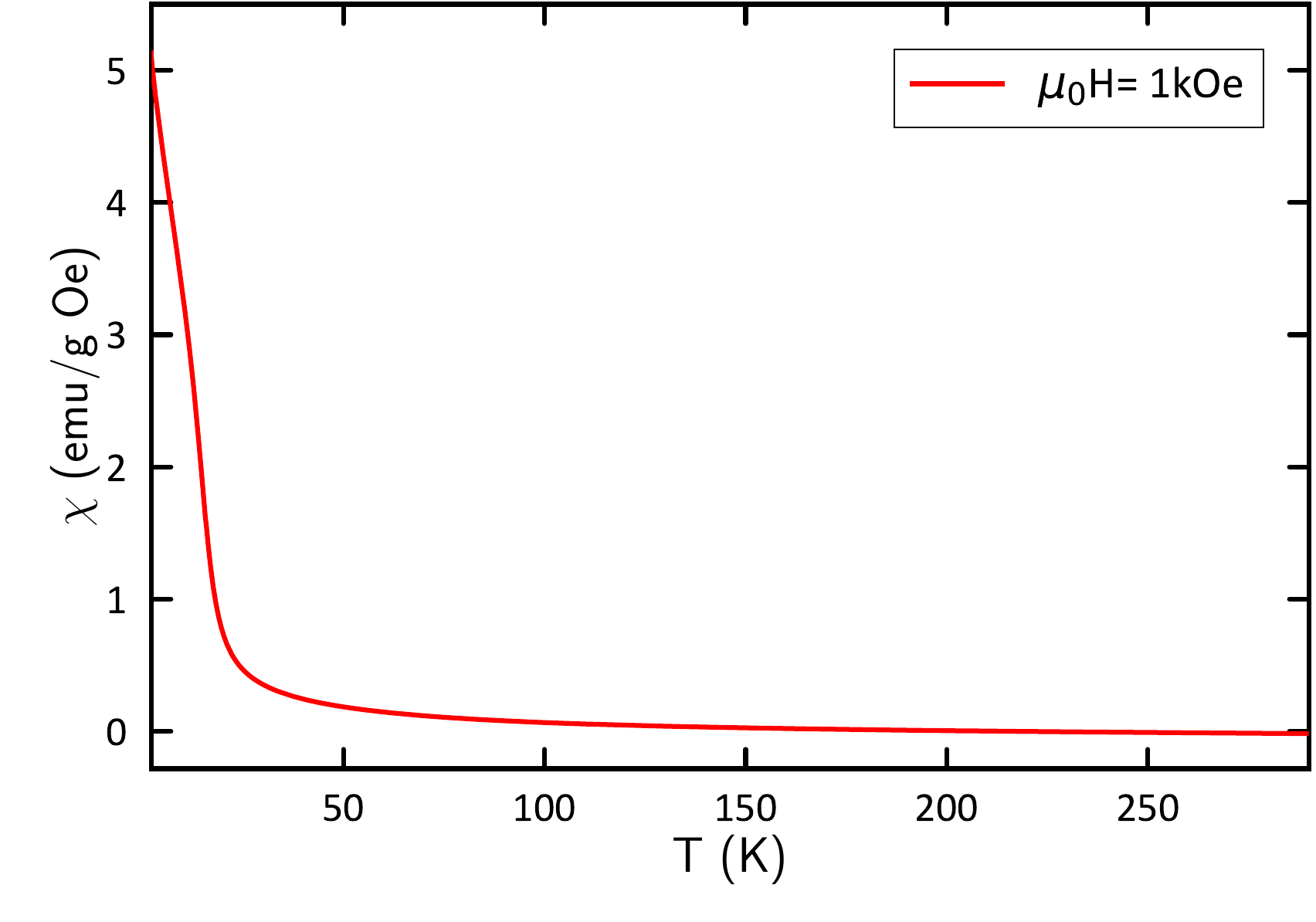}
    \caption{Temperature dependent DC magnetization of single crystal Yb-\iMAX sample at 0.01 T. The applied field is aligned along the \textit{c*} axis of the crystal.}
    \label{fig:Yb_MvsT}
\end{figure}

Fig.~\ref{fig:Yb_MvsT} shows $\chi$(T) of Yb-\iMAX\ measured at 1~kOe. Unlike Gd- and Dy-\iMAX\ phases, no maximum of $\chi$(T) is observed, and Yb-\iMAX\ exhibits a paramagnetic behavior in the measured temperature range of 1.9 – 300~K.

Therefore, there is a gradual decrease in the AFM transition temperature from $T_N\approx27K$ in Gd-\iMAX to $T_N\approx15K$ in Dy-\iMAX down to Yb-\iMAX\, which shows no transition temperature down to $\approx2~K$. It is possible to associate this behavior to a chemical pressure by the decreasing Rare Earth atomic radius and as a result to a change in the magnetic phase as predicted in other 2D magnets such as the transition metal phosphorus chalcogenides~\cite{Chittari2016} and recently reported in FePS$_3$~\cite{Coak2021,Pawbake2022}.
\begin{figure}
    \centering
    \includegraphics[width=0.9\linewidth]{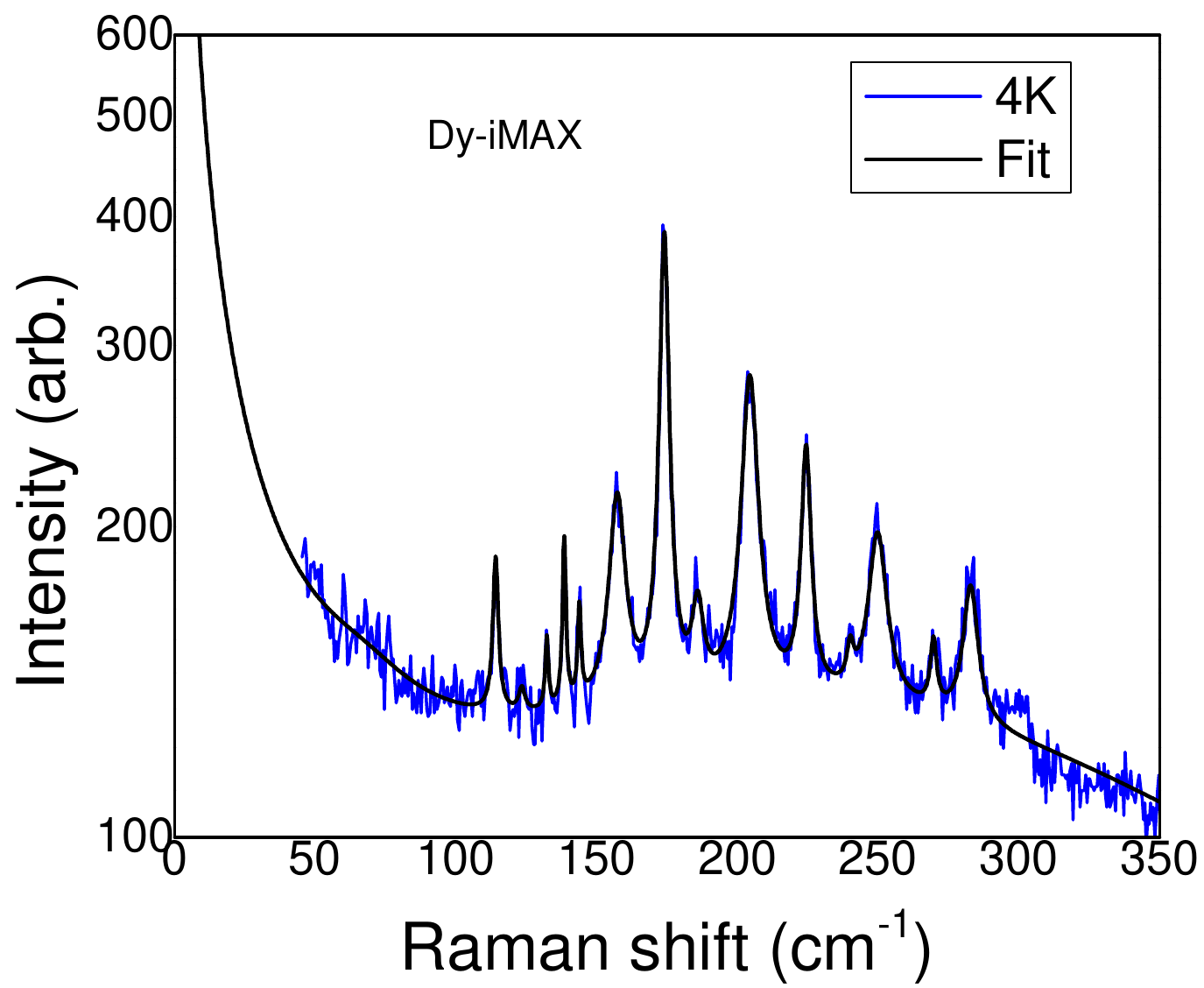}
    \caption{Raman spectra of Dy-\iMAX\ sample at 4 K. The
solid line shows the fitting to the data.}
    \label{fig:DyData4K}
\end{figure}

\begin{figure}
    \centering
    \includegraphics[width=0.9\linewidth]{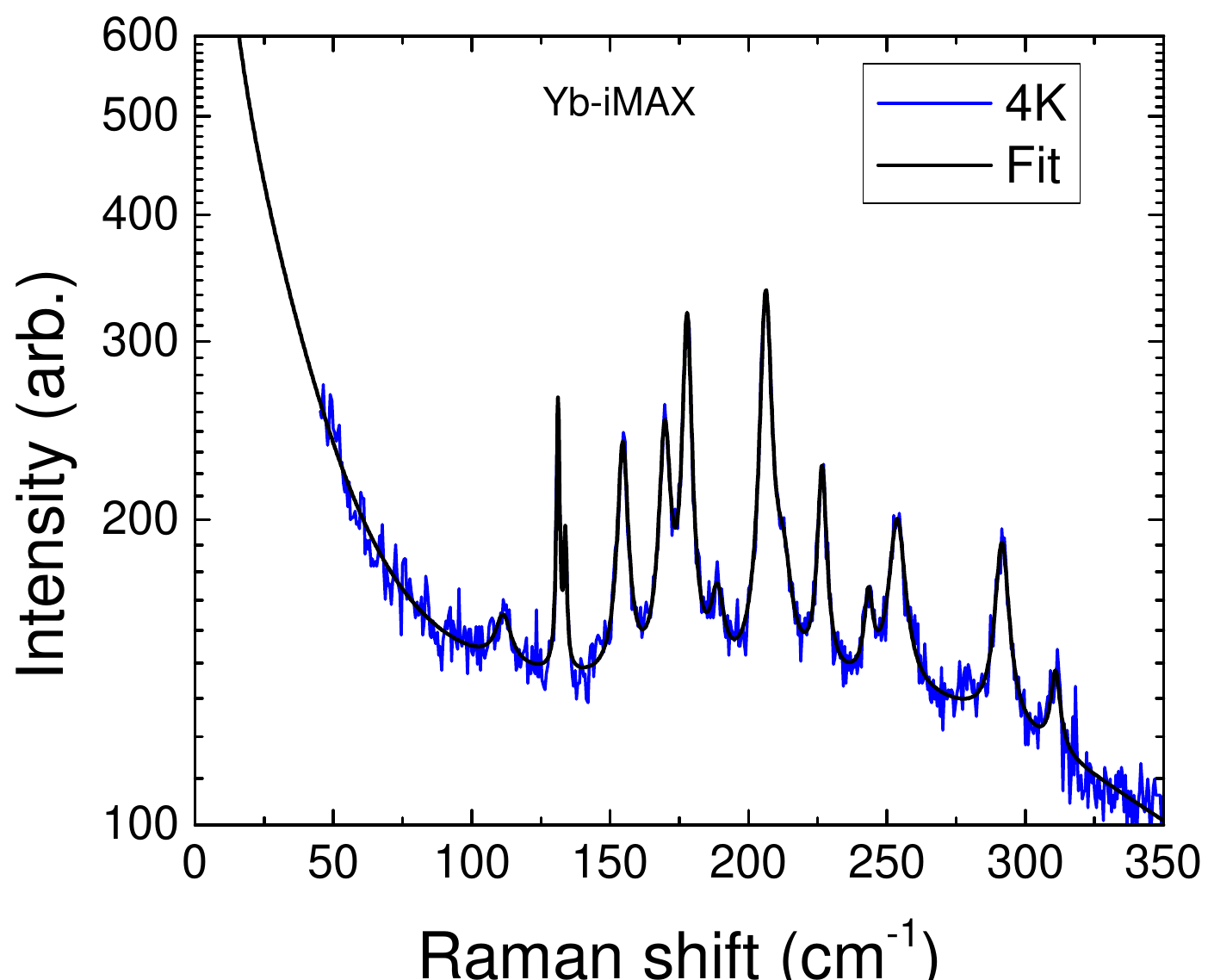}
    \caption{Raman spectra of Yb-\iMAX\ sample at 4 K. The
solid line shows the fitting to the data.}
    \label{fig:YbData4K}
\end{figure}

\section{Vibrational modes}

\begin{figure}
    \centering
    \includegraphics[width=0.75\linewidth]{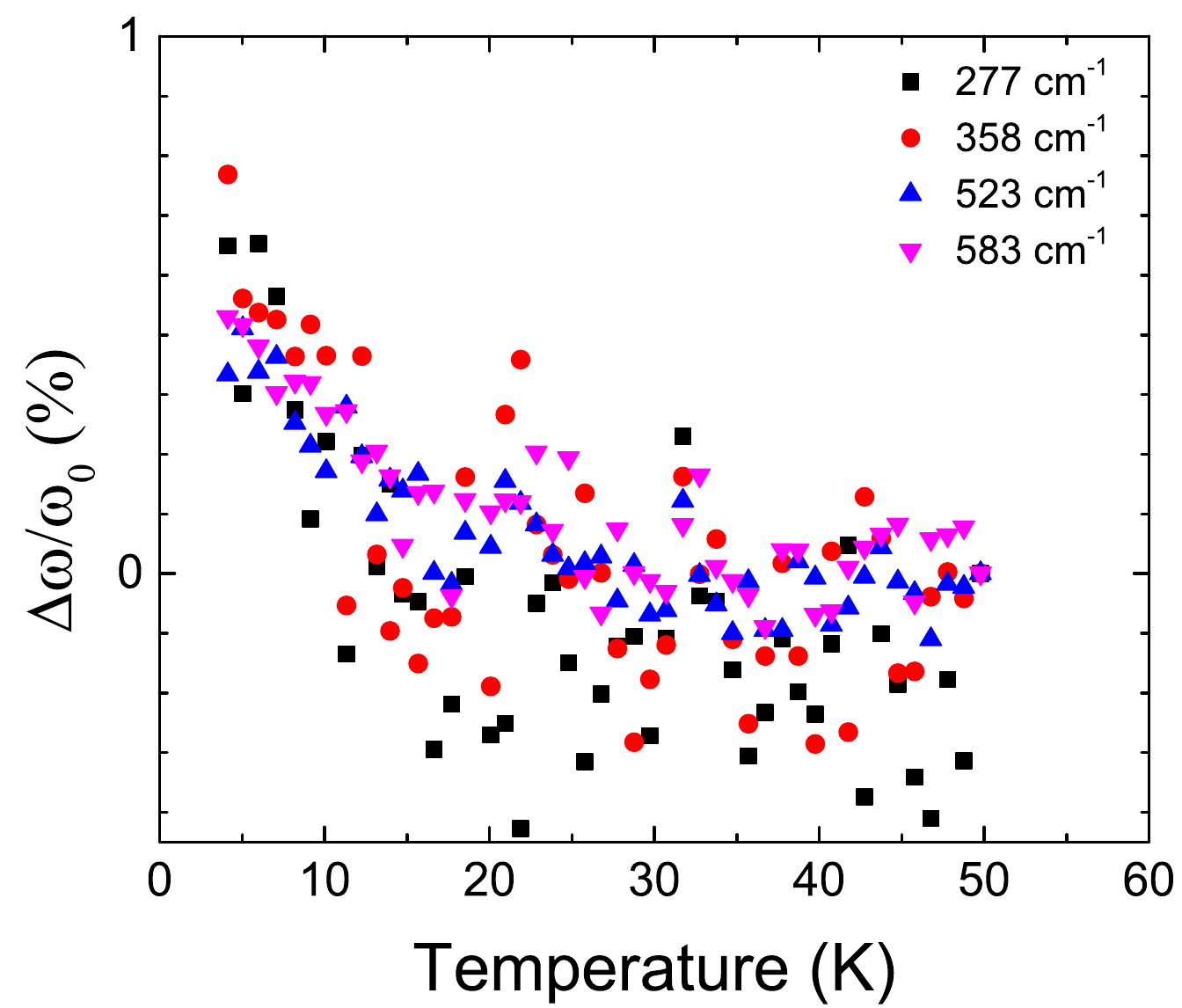}
    \caption{The relative change of the frequency normalized to the central frequency at base temperature as a function of temperature where $\Delta \omega = \omega-\omega_{0}$ for 277 cm$^{-1}$, 358 cm$^{-1}$,  523 cm$^{-1}$ and  583 cm$^{-1}$. }
    \label{fig:Gd_Omega0T_app}
\end{figure}

The hardening of the phonons below the antiferromagnetic phase transition seems less prominent in the high-energy phonons, as seen in Fig.~\ref{fig:Gd_Omega0T_app}. The change of less than 0.5\% in the normalized $\Delta \omega$ is of the order of the measurement error in the central frequency of the phonons above the phase transition.

The Raman shift spectrum of Dy- and Yb-\iMAX\ was measured at 4~K and fitted using RefFit in the same procedure as described above for the Gd-\iMAX\ sample. Figure~\ref{fig:DyData4K} and ~\ref{fig:YbData4K} shows the data (blue points) and the model (black line) for the Dy- and Yb-\iMAX\ samples, respectively. The exact values of the phonons' central frequencies as obtained from the fit are given in Table~\ref{tab:phonons} along with associated phonon symmetry and the theoretical values as calculated in Ref.~\citep{Champagne2019} for $Gd$- and $Dy$-\iMAX. Although there is no particular reasoning to compare the Raman intensities of the phonon modes, since these are different surfaces, their peak positions as a function of the chemical composition are of interest.

\begin{figure}
    \centering
    \includegraphics[width=0.9\linewidth]{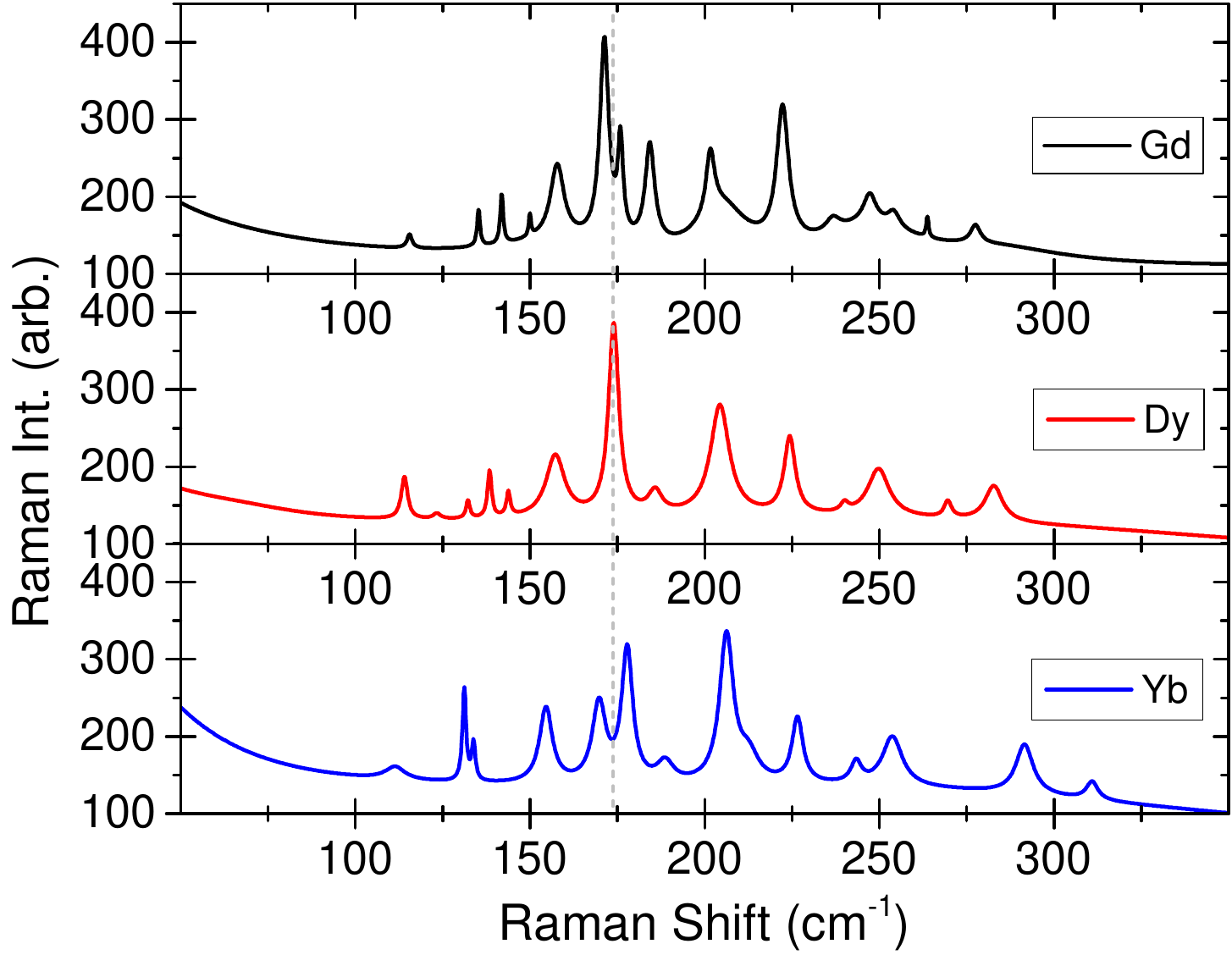}
    \caption{Raman spectra of Gd, Dy and Yb-\iMAX\ crystal at 4~K. The dashed line shows a guide to the eye of the central frequency of the strongest phonon at 171 cm$^{-1}$.}
    \label{fig:fitcomp}
\end{figure}

To quantitatively compare the evolution of the vibrational modes as a function of the Rare Earth atom, we plot in Figure~\ref{fig:fitcomp} the Raman model for each sample at 4~K and in the order of their appearance in the periodic table (and inverse order in their atomic radius). By comparing the different modes, it appears that the $B_g(1)$, $A_g(2)$, $B_g(3)$, $A_g(3)$, $A_g(4)$ phonon symmetries, which involve the Rare Earth atom, soften as the atomic radii reduce. On the other hand, the phonon modes $B_g(6)$, $B_g(7)$, $B_g(9)$, $A_g(9)$, $B_g(10)$, $B_g(13)$ which are mostly associated with the Mo and Al sublattices, seem to harden. In particular the $B_g(5)$ and $A_g(6)$ modes seem to show up as one peak in the $Dy$-\iMAX sample compared to two distinct peaks in the $Gd$- and $Yb$-\iMAX samples.   

\bibliography{refs}

\begin{thebibliography}{37}%
\makeatletter
\providecommand \@ifxundefined [1]{%
 \@ifx{#1\undefined}
}%
\providecommand \@ifnum [1]{%
 \ifnum #1\expandafter \@firstoftwo
 \else \expandafter \@secondoftwo
 \fi
}%
\providecommand \@ifx [1]{%
 \ifx #1\expandafter \@firstoftwo
 \else \expandafter \@secondoftwo
 \fi
}%
\providecommand \natexlab [1]{#1}%
\providecommand \enquote  [1]{``#1''}%
\providecommand \bibnamefont  [1]{#1}%
\providecommand \bibfnamefont [1]{#1}%
\providecommand \citenamefont [1]{#1}%
\providecommand \href@noop [0]{\@secondoftwo}%
\providecommand \href [0]{\begingroup \@sanitize@url \@href}%
\providecommand \@href[1]{\@@startlink{#1}\@@href}%
\providecommand \@@href[1]{\endgroup#1\@@endlink}%
\providecommand \@sanitize@url [0]{\catcode `\\12\catcode `\$12\catcode `\&12\catcode `\#12\catcode `\^12\catcode `\_12\catcode `\%12\relax}%
\providecommand \@@startlink[1]{}%
\providecommand \@@endlink[0]{}%
\providecommand \url  [0]{\begingroup\@sanitize@url \@url }%
\providecommand \@url [1]{\endgroup\@href {#1}{\urlprefix }}%
\providecommand \urlprefix  [0]{URL }%
\providecommand \Eprint [0]{\href }%
\providecommand \doibase [0]{https://doi.org/}%
\providecommand \selectlanguage [0]{\@gobble}%
\providecommand \bibinfo  [0]{\@secondoftwo}%
\providecommand \bibfield  [0]{\@secondoftwo}%
\providecommand \translation [1]{[#1]}%
\providecommand \BibitemOpen [0]{}%
\providecommand \bibitemStop [0]{}%
\providecommand \bibitemNoStop [0]{.\EOS\space}%
\providecommand \EOS [0]{\spacefactor3000\relax}%
\providecommand \BibitemShut  [1]{\csname bibitem#1\endcsname}%
\let\auto@bib@innerbib\@empty
\bibitem [{\citenamefont {Lockwood}\ and\ \citenamefont {Cottam}(1988)}]{Lockwood1988}%
  \BibitemOpen
  \bibfield  {author} {\bibinfo {author} {\bibfnamefont {D.~J.}\ \bibnamefont {Lockwood}}\ and\ \bibinfo {author} {\bibfnamefont {M.~G.}\ \bibnamefont {Cottam}},\ }\bibfield  {title} {\bibinfo {title} {{The spin-phonon interaction in FeF$_{2}$ and MnF$_{2}$ studied by Raman spectroscopy}},\ }\href {https://doi.org/10.1063/1.342186} {\bibfield  {journal} {\bibinfo  {journal} {Journal of Applied Physics}\ }\textbf {\bibinfo {volume} {64}},\ \bibinfo {pages} {5876–5878} (\bibinfo {year} {1988})}\BibitemShut {NoStop}%
\bibitem [{\citenamefont {Huang}\ \emph {et~al.}(2017)\citenamefont {Huang}, \citenamefont {Clark}, \citenamefont {Navarro-Moratalla}, \citenamefont {Klein}, \citenamefont {Cheng}, \citenamefont {Seyler}, \citenamefont {Zhong}, \citenamefont {Schmidgall}, \citenamefont {McGuire}, \citenamefont {Cobden}, \citenamefont {Yao}, \citenamefont {Xiao}, \citenamefont {Jarillo-Herrero},\ and\ \citenamefont {Xu}}]{Huang2017}%
  \BibitemOpen
  \bibfield  {author} {\bibinfo {author} {\bibfnamefont {B.}~\bibnamefont {Huang}}, \bibinfo {author} {\bibfnamefont {G.}~\bibnamefont {Clark}}, \bibinfo {author} {\bibfnamefont {E.}~\bibnamefont {Navarro-Moratalla}}, \bibinfo {author} {\bibfnamefont {D.~R.}\ \bibnamefont {Klein}}, \bibinfo {author} {\bibfnamefont {R.}~\bibnamefont {Cheng}}, \bibinfo {author} {\bibfnamefont {K.~L.}\ \bibnamefont {Seyler}}, \bibinfo {author} {\bibfnamefont {D.}~\bibnamefont {Zhong}}, \bibinfo {author} {\bibfnamefont {E.}~\bibnamefont {Schmidgall}}, \bibinfo {author} {\bibfnamefont {M.~A.}\ \bibnamefont {McGuire}}, \bibinfo {author} {\bibfnamefont {D.~H.}\ \bibnamefont {Cobden}}, \bibinfo {author} {\bibfnamefont {W.}~\bibnamefont {Yao}}, \bibinfo {author} {\bibfnamefont {D.}~\bibnamefont {Xiao}}, \bibinfo {author} {\bibfnamefont {P.}~\bibnamefont {Jarillo-Herrero}},\ and\ \bibinfo {author} {\bibfnamefont {X.}~\bibnamefont {Xu}},\ }\bibfield  {title} {\bibinfo {title} {{Layer-dependent ferromagnetism in a van der Waals crystal
  down to the monolayer limit}},\ }\href {https://doi.org/10.1038/nature22391} {\bibfield  {journal} {\bibinfo  {journal} {Nature}\ }\textbf {\bibinfo {volume} {546}},\ \bibinfo {pages} {270–273} (\bibinfo {year} {2017})}\BibitemShut {NoStop}%
\bibitem [{\citenamefont {Fransson}\ \emph {et~al.}(2016)\citenamefont {Fransson}, \citenamefont {Black-Schaffer},\ and\ \citenamefont {Balatsky}}]{fransson2016magnon}%
  \BibitemOpen
  \bibfield  {author} {\bibinfo {author} {\bibfnamefont {J.}~\bibnamefont {Fransson}}, \bibinfo {author} {\bibfnamefont {A.~M.}\ \bibnamefont {Black-Schaffer}},\ and\ \bibinfo {author} {\bibfnamefont {A.~V.}\ \bibnamefont {Balatsky}},\ }\bibfield  {title} {\bibinfo {title} {{Magnon Dirac materials}},\ }\href {https://doi.org/10.1103/physrevb.94.075401} {\bibfield  {journal} {\bibinfo  {journal} {Physical Review B}\ }\textbf {\bibinfo {volume} {94}},\ \bibinfo {pages} {075401} (\bibinfo {year} {2016})}\BibitemShut {NoStop}%
\bibitem [{\citenamefont {Pershoguba}\ \emph {et~al.}(2018)\citenamefont {Pershoguba}, \citenamefont {Banerjee}, \citenamefont {Lashley}, \citenamefont {Park}, \citenamefont {Ågren}, \citenamefont {Aeppli},\ and\ \citenamefont {Balatsky}}]{pershoguba2018dirac}%
  \BibitemOpen
  \bibfield  {author} {\bibinfo {author} {\bibfnamefont {S.~S.}\ \bibnamefont {Pershoguba}}, \bibinfo {author} {\bibfnamefont {S.}~\bibnamefont {Banerjee}}, \bibinfo {author} {\bibfnamefont {J.}~\bibnamefont {Lashley}}, \bibinfo {author} {\bibfnamefont {J.}~\bibnamefont {Park}}, \bibinfo {author} {\bibfnamefont {H.}~\bibnamefont {Ågren}}, \bibinfo {author} {\bibfnamefont {G.}~\bibnamefont {Aeppli}},\ and\ \bibinfo {author} {\bibfnamefont {A.~V.}\ \bibnamefont {Balatsky}},\ }\bibfield  {title} {\bibinfo {title} {{Dirac Magnons in Honeycomb Ferromagnets}},\ }\href {https://doi.org/10.1103/physrevx.8.011010} {\bibfield  {journal} {\bibinfo  {journal} {Physical Review X}\ }\textbf {\bibinfo {volume} {8}},\ \bibinfo {pages} {011010} (\bibinfo {year} {2018})}\BibitemShut {NoStop}%
\bibitem [{\citenamefont {Chen}\ \emph {et~al.}(2018)\citenamefont {Chen}, \citenamefont {Chung}, \citenamefont {Gao}, \citenamefont {Chen}, \citenamefont {Stone}, \citenamefont {Kolesnikov}, \citenamefont {Huang},\ and\ \citenamefont {Dai}}]{chen2018topological}%
  \BibitemOpen
  \bibfield  {author} {\bibinfo {author} {\bibfnamefont {L.}~\bibnamefont {Chen}}, \bibinfo {author} {\bibfnamefont {J.-H.}\ \bibnamefont {Chung}}, \bibinfo {author} {\bibfnamefont {B.}~\bibnamefont {Gao}}, \bibinfo {author} {\bibfnamefont {T.}~\bibnamefont {Chen}}, \bibinfo {author} {\bibfnamefont {M.~B.}\ \bibnamefont {Stone}}, \bibinfo {author} {\bibfnamefont {A.~I.}\ \bibnamefont {Kolesnikov}}, \bibinfo {author} {\bibfnamefont {Q.}~\bibnamefont {Huang}},\ and\ \bibinfo {author} {\bibfnamefont {P.}~\bibnamefont {Dai}},\ }\bibfield  {title} {\bibinfo {title} {{Topological Spin Excitations in Honeycomb Ferromagnet ${\mathrm{CrI}}_{3}$}},\ }\href {https://doi.org/10.1103/PhysRevX.8.041028} {\bibfield  {journal} {\bibinfo  {journal} {Phys. Rev. X}\ }\textbf {\bibinfo {volume} {8}},\ \bibinfo {pages} {041028} (\bibinfo {year} {2018})}\BibitemShut {NoStop}%
\bibitem [{\citenamefont {Chen}\ \emph {et~al.}(2021)\citenamefont {Chen}, \citenamefont {Chung}, \citenamefont {Stone}, \citenamefont {Kolesnikov}, \citenamefont {Winn}, \citenamefont {Garlea}, \citenamefont {Abernathy}, \citenamefont {Gao}, \citenamefont {Augustin}, \citenamefont {Santos},\ and\ \citenamefont {Dai}}]{chen2021magnetic}%
  \BibitemOpen
  \bibfield  {author} {\bibinfo {author} {\bibfnamefont {L.}~\bibnamefont {Chen}}, \bibinfo {author} {\bibfnamefont {J.-H.}\ \bibnamefont {Chung}}, \bibinfo {author} {\bibfnamefont {M.~B.}\ \bibnamefont {Stone}}, \bibinfo {author} {\bibfnamefont {A.~I.}\ \bibnamefont {Kolesnikov}}, \bibinfo {author} {\bibfnamefont {B.}~\bibnamefont {Winn}}, \bibinfo {author} {\bibfnamefont {V.~O.}\ \bibnamefont {Garlea}}, \bibinfo {author} {\bibfnamefont {D.~L.}\ \bibnamefont {Abernathy}}, \bibinfo {author} {\bibfnamefont {B.}~\bibnamefont {Gao}}, \bibinfo {author} {\bibfnamefont {M.}~\bibnamefont {Augustin}}, \bibinfo {author} {\bibfnamefont {E.~J.~G.}\ \bibnamefont {Santos}},\ and\ \bibinfo {author} {\bibfnamefont {P.}~\bibnamefont {Dai}},\ }\bibfield  {title} {\bibinfo {title} {{Magnetic Field Effect on Topological Spin Excitations in ${\mathrm{CrI}}_{3}$}},\ }\href {https://doi.org/10.1103/PhysRevX.11.031047} {\bibfield  {journal} {\bibinfo  {journal} {Phys. Rev. X}\ }\textbf {\bibinfo {volume} {11}},\ \bibinfo {pages}
  {031047} (\bibinfo {year} {2021})}\BibitemShut {NoStop}%
\bibitem [{\citenamefont {Sun}\ \emph {et~al.}(2023)\citenamefont {Sun}, \citenamefont {Bhowmick}, \citenamefont {Yang},\ and\ \citenamefont {Sengupta}}]{sun2023interacting}%
  \BibitemOpen
  \bibfield  {author} {\bibinfo {author} {\bibfnamefont {H.}~\bibnamefont {Sun}}, \bibinfo {author} {\bibfnamefont {D.}~\bibnamefont {Bhowmick}}, \bibinfo {author} {\bibfnamefont {B.}~\bibnamefont {Yang}},\ and\ \bibinfo {author} {\bibfnamefont {P.}~\bibnamefont {Sengupta}},\ }\bibfield  {title} {\bibinfo {title} {{Interacting topological Dirac magnons}},\ }\href {https://doi.org/10.1103/physrevb.107.134426} {\bibfield  {journal} {\bibinfo  {journal} {Physical Review B}\ }\textbf {\bibinfo {volume} {107}},\ \bibinfo {pages} {134426} (\bibinfo {year} {2023})}\BibitemShut {NoStop}%
\bibitem [{\citenamefont {Klein}\ \emph {et~al.}(2019)\citenamefont {Klein}, \citenamefont {MacNeill}, \citenamefont {Song}, \citenamefont {Larson}, \citenamefont {Fang}, \citenamefont {Xu}, \citenamefont {Ribeiro}, \citenamefont {Canfield}, \citenamefont {Kaxiras}, \citenamefont {Comin},\ and\ \citenamefont {Jarillo-Herrero}}]{klein2019enhancement}%
  \BibitemOpen
  \bibfield  {author} {\bibinfo {author} {\bibfnamefont {D.~R.}\ \bibnamefont {Klein}}, \bibinfo {author} {\bibfnamefont {D.}~\bibnamefont {MacNeill}}, \bibinfo {author} {\bibfnamefont {Q.}~\bibnamefont {Song}}, \bibinfo {author} {\bibfnamefont {D.~T.}\ \bibnamefont {Larson}}, \bibinfo {author} {\bibfnamefont {S.}~\bibnamefont {Fang}}, \bibinfo {author} {\bibfnamefont {M.}~\bibnamefont {Xu}}, \bibinfo {author} {\bibfnamefont {R.~A.}\ \bibnamefont {Ribeiro}}, \bibinfo {author} {\bibfnamefont {P.~C.}\ \bibnamefont {Canfield}}, \bibinfo {author} {\bibfnamefont {E.}~\bibnamefont {Kaxiras}}, \bibinfo {author} {\bibfnamefont {R.}~\bibnamefont {Comin}},\ and\ \bibinfo {author} {\bibfnamefont {P.}~\bibnamefont {Jarillo-Herrero}},\ }\bibfield  {title} {\bibinfo {title} {{Enhancement of interlayer exchange in an ultrathin two-dimensional magnet}},\ }\href {https://doi.org/10.1038/s41567-019-0651-0} {\bibfield  {journal} {\bibinfo  {journal} {Nature Physics}\ }\textbf {\bibinfo {volume} {15}},\ \bibinfo {pages}
  {1255–1260} (\bibinfo {year} {2019})}\BibitemShut {NoStop}%
\bibitem [{\citenamefont {Zhong}\ \emph {et~al.}(2017)\citenamefont {Zhong}, \citenamefont {Seyler}, \citenamefont {Linpeng}, \citenamefont {Cheng}, \citenamefont {Sivadas}, \citenamefont {Huang}, \citenamefont {Schmidgall}, \citenamefont {Taniguchi}, \citenamefont {Watanabe}, \citenamefont {McGuire}, \citenamefont {Yao}, \citenamefont {Xiao}, \citenamefont {Fu},\ and\ \citenamefont {Xu}}]{zhong2017van}%
  \BibitemOpen
  \bibfield  {author} {\bibinfo {author} {\bibfnamefont {D.}~\bibnamefont {Zhong}}, \bibinfo {author} {\bibfnamefont {K.~L.}\ \bibnamefont {Seyler}}, \bibinfo {author} {\bibfnamefont {X.}~\bibnamefont {Linpeng}}, \bibinfo {author} {\bibfnamefont {R.}~\bibnamefont {Cheng}}, \bibinfo {author} {\bibfnamefont {N.}~\bibnamefont {Sivadas}}, \bibinfo {author} {\bibfnamefont {B.}~\bibnamefont {Huang}}, \bibinfo {author} {\bibfnamefont {E.}~\bibnamefont {Schmidgall}}, \bibinfo {author} {\bibfnamefont {T.}~\bibnamefont {Taniguchi}}, \bibinfo {author} {\bibfnamefont {K.}~\bibnamefont {Watanabe}}, \bibinfo {author} {\bibfnamefont {M.~A.}\ \bibnamefont {McGuire}}, \bibinfo {author} {\bibfnamefont {W.}~\bibnamefont {Yao}}, \bibinfo {author} {\bibfnamefont {D.}~\bibnamefont {Xiao}}, \bibinfo {author} {\bibfnamefont {K.-M.~C.}\ \bibnamefont {Fu}},\ and\ \bibinfo {author} {\bibfnamefont {X.}~\bibnamefont {Xu}},\ }\bibfield  {title} {\bibinfo {title} {{Van der Waals engineering of ferromagnetic semiconductor heterostructures
  for spin and valleytronics}},\ }\href {https://doi.org/10.1126/sciadv.1603113} {\bibfield  {journal} {\bibinfo  {journal} {Science Advances}\ }\textbf {\bibinfo {volume} {3}},\ \bibinfo {pages} {e1603113} (\bibinfo {year} {2017})}\BibitemShut {NoStop}%
\bibitem [{\citenamefont {Ziebel}\ \emph {et~al.}(2024)\citenamefont {Ziebel}, \citenamefont {Feuer}, \citenamefont {Cox}, \citenamefont {Zhu}, \citenamefont {Dean},\ and\ \citenamefont {Roy}}]{ziebel2024crsbr}%
  \BibitemOpen
  \bibfield  {author} {\bibinfo {author} {\bibfnamefont {M.~E.}\ \bibnamefont {Ziebel}}, \bibinfo {author} {\bibfnamefont {M.~L.}\ \bibnamefont {Feuer}}, \bibinfo {author} {\bibfnamefont {J.}~\bibnamefont {Cox}}, \bibinfo {author} {\bibfnamefont {X.}~\bibnamefont {Zhu}}, \bibinfo {author} {\bibfnamefont {C.~R.}\ \bibnamefont {Dean}},\ and\ \bibinfo {author} {\bibfnamefont {X.}~\bibnamefont {Roy}},\ }\bibfield  {title} {\bibinfo {title} {{CrSBr}: {An} {Air}-{Stable}, {Two}-{Dimensional} {Magnetic} {Semiconductor}},\ }\href {https://doi.org/10.1021/acs.nanolett.4c00624} {\bibfield  {journal} {\bibinfo  {journal} {Nano Lett.}\ }\textbf {\bibinfo {volume} {24}},\ \bibinfo {pages} {4319} (\bibinfo {year} {2024})},\ \bibinfo {note} {publisher: American Chemical Society}\BibitemShut {NoStop}%
\bibitem [{\citenamefont {Yang}\ \emph {et~al.}(2020)\citenamefont {Yang}, \citenamefont {Zhang},\ and\ \citenamefont {Jiang}}]{Yang2020}%
  \BibitemOpen
  \bibfield  {author} {\bibinfo {author} {\bibfnamefont {S.}~\bibnamefont {Yang}}, \bibinfo {author} {\bibfnamefont {T.}~\bibnamefont {Zhang}},\ and\ \bibinfo {author} {\bibfnamefont {C.}~\bibnamefont {Jiang}},\ }\bibfield  {title} {\bibinfo {title} {{van der Waals Magnets: Material Family, Detection and Modulation of Magnetism, and Perspective in Spintronics}},\ }\href {https://doi.org/10.1002/advs.202002488} {\bibfield  {journal} {\bibinfo  {journal} {Advanced Science}\ }\textbf {\bibinfo {volume} {8}},\ \bibinfo {pages} {2002488} (\bibinfo {year} {2020})}\BibitemShut {NoStop}%
\bibitem [{\citenamefont {Gibertini}\ \emph {et~al.}(2019)\citenamefont {Gibertini}, \citenamefont {Koperski}, \citenamefont {Morpurgo},\ and\ \citenamefont {Novoselov}}]{Gibertini2019}%
  \BibitemOpen
  \bibfield  {author} {\bibinfo {author} {\bibfnamefont {M.}~\bibnamefont {Gibertini}}, \bibinfo {author} {\bibfnamefont {M.}~\bibnamefont {Koperski}}, \bibinfo {author} {\bibfnamefont {A.~F.}\ \bibnamefont {Morpurgo}},\ and\ \bibinfo {author} {\bibfnamefont {K.~S.}\ \bibnamefont {Novoselov}},\ }\bibfield  {title} {\bibinfo {title} {{Magnetic 2D materials and heterostructures}},\ }\href {https://doi.org/10.1038/s41565-019-0438-6} {\bibfield  {journal} {\bibinfo  {journal} {Nature Nanotechnology}\ }\textbf {\bibinfo {volume} {14}},\ \bibinfo {pages} {408–419} (\bibinfo {year} {2019})}\BibitemShut {NoStop}%
\bibitem [{\citenamefont {Wang}\ \emph {et~al.}(2016)\citenamefont {Wang}, \citenamefont {Du}, \citenamefont {Fredrik~Liu}, \citenamefont {Hu}, \citenamefont {Zhang}, \citenamefont {Zhang}, \citenamefont {Owen}, \citenamefont {Lu}, \citenamefont {Gan}, \citenamefont {Sengupta}, \citenamefont {Kloc},\ and\ \citenamefont {Xiong}}]{Wang2016}%
  \BibitemOpen
  \bibfield  {author} {\bibinfo {author} {\bibfnamefont {X.}~\bibnamefont {Wang}}, \bibinfo {author} {\bibfnamefont {K.}~\bibnamefont {Du}}, \bibinfo {author} {\bibfnamefont {Y.~Y.}\ \bibnamefont {Fredrik~Liu}}, \bibinfo {author} {\bibfnamefont {P.}~\bibnamefont {Hu}}, \bibinfo {author} {\bibfnamefont {J.}~\bibnamefont {Zhang}}, \bibinfo {author} {\bibfnamefont {Q.}~\bibnamefont {Zhang}}, \bibinfo {author} {\bibfnamefont {M.~H.~S.}\ \bibnamefont {Owen}}, \bibinfo {author} {\bibfnamefont {X.}~\bibnamefont {Lu}}, \bibinfo {author} {\bibfnamefont {C.~K.}\ \bibnamefont {Gan}}, \bibinfo {author} {\bibfnamefont {P.}~\bibnamefont {Sengupta}}, \bibinfo {author} {\bibfnamefont {C.}~\bibnamefont {Kloc}},\ and\ \bibinfo {author} {\bibfnamefont {Q.}~\bibnamefont {Xiong}},\ }\bibfield  {title} {\bibinfo {title} {{Raman spectroscopy of atomically thin two-dimensional magnetic iron phosphorus trisulfide {FePS$_{3}$} crystals}},\ }\href {https://doi.org/10.1088/2053-1583/3/3/031009} {\bibfield  {journal} {\bibinfo  {journal}
  {2D Materials}\ }\textbf {\bibinfo {volume} {3}},\ \bibinfo {pages} {031009} (\bibinfo {year} {2016})}\BibitemShut {NoStop}%
\bibitem [{\citenamefont {Lee}\ \emph {et~al.}(2016)\citenamefont {Lee}, \citenamefont {Lee}, \citenamefont {Ryoo}, \citenamefont {Kang}, \citenamefont {Kim}, \citenamefont {Kim}, \citenamefont {Park}, \citenamefont {Park},\ and\ \citenamefont {Cheong}}]{Lee2016}%
  \BibitemOpen
  \bibfield  {author} {\bibinfo {author} {\bibfnamefont {J.-U.}\ \bibnamefont {Lee}}, \bibinfo {author} {\bibfnamefont {S.}~\bibnamefont {Lee}}, \bibinfo {author} {\bibfnamefont {J.~H.}\ \bibnamefont {Ryoo}}, \bibinfo {author} {\bibfnamefont {S.}~\bibnamefont {Kang}}, \bibinfo {author} {\bibfnamefont {T.~Y.}\ \bibnamefont {Kim}}, \bibinfo {author} {\bibfnamefont {P.}~\bibnamefont {Kim}}, \bibinfo {author} {\bibfnamefont {C.-H.}\ \bibnamefont {Park}}, \bibinfo {author} {\bibfnamefont {J.-G.}\ \bibnamefont {Park}},\ and\ \bibinfo {author} {\bibfnamefont {H.}~\bibnamefont {Cheong}},\ }\bibfield  {title} {\bibinfo {title} {{Ising-Type Magnetic Ordering in Atomically Thin {FePS$_{3}$}}},\ }\href {https://doi.org/10.1021/acs.nanolett.6b03052} {\bibfield  {journal} {\bibinfo  {journal} {Nano Letters}\ }\textbf {\bibinfo {volume} {16}},\ \bibinfo {pages} {7433–7438} (\bibinfo {year} {2016})}\BibitemShut {NoStop}%
\bibitem [{\citenamefont {Petruhins}\ \emph {et~al.}(2019)\citenamefont {Petruhins}, \citenamefont {Lu}, \citenamefont {Hultman},\ and\ \citenamefont {Rosen}}]{Petruhins2019}%
  \BibitemOpen
  \bibfield  {author} {\bibinfo {author} {\bibfnamefont {A.}~\bibnamefont {Petruhins}}, \bibinfo {author} {\bibfnamefont {J.}~\bibnamefont {Lu}}, \bibinfo {author} {\bibfnamefont {L.}~\bibnamefont {Hultman}},\ and\ \bibinfo {author} {\bibfnamefont {J.}~\bibnamefont {Rosen}},\ }\bibfield  {title} {\bibinfo {title} {{Synthesis of atomically layered and chemically ordered rare-earth (RE) i-MAX phases; (Mo2/3RE1/3)2GaC with RE = Gd, Tb, Dy, Ho, Er, Tm, Yb, and Lu}},\ }\href {https://doi.org/10.1080/21663831.2019.1644684} {\bibfield  {journal} {\bibinfo  {journal} {Materials Research Letters}\ }\textbf {\bibinfo {volume} {7}},\ \bibinfo {pages} {446–452} (\bibinfo {year} {2019})}\BibitemShut {NoStop}%
\bibitem [{\citenamefont {Dahlqvist}\ and\ \citenamefont {Rosen}(2022)}]{Dahlqvist2022}%
  \BibitemOpen
  \bibfield  {author} {\bibinfo {author} {\bibfnamefont {M.}~\bibnamefont {Dahlqvist}}\ and\ \bibinfo {author} {\bibfnamefont {J.}~\bibnamefont {Rosen}},\ }\bibfield  {title} {\bibinfo {title} {{The rise of MAX phase alloys – large-scale theoretical screening for the prediction of chemical order and disorder}},\ }\href {https://doi.org/10.1039/d2nr02414d} {\bibfield  {journal} {\bibinfo  {journal} {Nanoscale}\ }\textbf {\bibinfo {volume} {14}},\ \bibinfo {pages} {10958–10971} (\bibinfo {year} {2022})}\BibitemShut {NoStop}%
\bibitem [{\citenamefont {Tao}\ \emph {et~al.}(2017)\citenamefont {Tao}, \citenamefont {Dahlqvist}, \citenamefont {Lu}, \citenamefont {Kota}, \citenamefont {Meshkian}, \citenamefont {Halim}, \citenamefont {Palisaitis}, \citenamefont {Hultman}, \citenamefont {Barsoum}, \citenamefont {Persson},\ and\ \citenamefont {Rosen}}]{tao2017two}%
  \BibitemOpen
  \bibfield  {author} {\bibinfo {author} {\bibfnamefont {Q.}~\bibnamefont {Tao}}, \bibinfo {author} {\bibfnamefont {M.}~\bibnamefont {Dahlqvist}}, \bibinfo {author} {\bibfnamefont {J.}~\bibnamefont {Lu}}, \bibinfo {author} {\bibfnamefont {S.}~\bibnamefont {Kota}}, \bibinfo {author} {\bibfnamefont {R.}~\bibnamefont {Meshkian}}, \bibinfo {author} {\bibfnamefont {J.}~\bibnamefont {Halim}}, \bibinfo {author} {\bibfnamefont {J.}~\bibnamefont {Palisaitis}}, \bibinfo {author} {\bibfnamefont {L.}~\bibnamefont {Hultman}}, \bibinfo {author} {\bibfnamefont {M.~W.}\ \bibnamefont {Barsoum}}, \bibinfo {author} {\bibfnamefont {P.~O.}\ \bibnamefont {Persson}},\ and\ \bibinfo {author} {\bibfnamefont {J.}~\bibnamefont {Rosen}},\ }\bibfield  {title} {\bibinfo {title} {{Two-dimensional Mo$_{1.33}$C MXene with divacancy ordering prepared from parent 3D laminate with in-plane chemical ordering}},\ }\href {https://doi.org/10.1038/ncomms14949} {\bibfield  {journal} {\bibinfo  {journal} {Nature Communications}\ }\textbf {\bibinfo
  {volume} {8}},\ \bibinfo {pages} {14949} (\bibinfo {year} {2017})}\BibitemShut {NoStop}%
\bibitem [{\citenamefont {Ohlmann}\ and\ \citenamefont {Tinkham}(1961)}]{Ohlmann1961}%
  \BibitemOpen
  \bibfield  {author} {\bibinfo {author} {\bibfnamefont {R.~C.}\ \bibnamefont {Ohlmann}}\ and\ \bibinfo {author} {\bibfnamefont {M.}~\bibnamefont {Tinkham}},\ }\bibfield  {title} {\bibinfo {title} {{Antiferromagnetic Resonance in Fe${\mathrm{F}}_{2}$ at Far-Infrared Frequencies}},\ }\href {https://doi.org/10.1103/PhysRev.123.425} {\bibfield  {journal} {\bibinfo  {journal} {Phys. Rev.}\ }\textbf {\bibinfo {volume} {123}},\ \bibinfo {pages} {425} (\bibinfo {year} {1961})}\BibitemShut {NoStop}%
\bibitem [{\citenamefont {Talbayev}\ \emph {et~al.}(2004)\citenamefont {Talbayev}, \citenamefont {Mih\'aly},\ and\ \citenamefont {Zhou}}]{Talbayev2004}%
  \BibitemOpen
  \bibfield  {author} {\bibinfo {author} {\bibfnamefont {D.}~\bibnamefont {Talbayev}}, \bibinfo {author} {\bibfnamefont {L.}~\bibnamefont {Mih\'aly}},\ and\ \bibinfo {author} {\bibfnamefont {J.}~\bibnamefont {Zhou}},\ }\bibfield  {title} {\bibinfo {title} {{Antiferromagnetic Resonance in ${\mathrm{L}\mathrm{a}\mathrm{M}\mathrm{n}\mathrm{O}}_{3}$ at Low Temperature}},\ }\href {https://doi.org/10.1103/PhysRevLett.93.017202} {\bibfield  {journal} {\bibinfo  {journal} {Phys. Rev. Lett.}\ }\textbf {\bibinfo {volume} {93}},\ \bibinfo {pages} {017202} (\bibinfo {year} {2004})}\BibitemShut {NoStop}%
\bibitem [{\citenamefont {Wildes}\ \emph {et~al.}(2012)\citenamefont {Wildes}, \citenamefont {Rule}, \citenamefont {Bewley}, \citenamefont {Enderle},\ and\ \citenamefont {Hicks}}]{Wildes2012}%
  \BibitemOpen
  \bibfield  {author} {\bibinfo {author} {\bibfnamefont {A.~R.}\ \bibnamefont {Wildes}}, \bibinfo {author} {\bibfnamefont {K.~C.}\ \bibnamefont {Rule}}, \bibinfo {author} {\bibfnamefont {R.~I.}\ \bibnamefont {Bewley}}, \bibinfo {author} {\bibfnamefont {M.}~\bibnamefont {Enderle}},\ and\ \bibinfo {author} {\bibfnamefont {T.~J.}\ \bibnamefont {Hicks}},\ }\bibfield  {title} {\bibinfo {title} {{The magnon dynamics and spin exchange parameters of FePS$_{3}$}},\ }\href {https://doi.org/10.1088/0953-8984/24/41/416004} {\bibfield  {journal} {\bibinfo  {journal} {Journal of Physics: Condensed Matter}\ }\textbf {\bibinfo {volume} {24}},\ \bibinfo {pages} {416004} (\bibinfo {year} {2012})}\BibitemShut {NoStop}%
\bibitem [{\citenamefont {Basov}\ \emph {et~al.}(2011)\citenamefont {Basov}, \citenamefont {Averitt}, \citenamefont {van~der Marel}, \citenamefont {Dressel},\ and\ \citenamefont {Haule}}]{Basov2011}%
  \BibitemOpen
  \bibfield  {author} {\bibinfo {author} {\bibfnamefont {D.~N.}\ \bibnamefont {Basov}}, \bibinfo {author} {\bibfnamefont {R.~D.}\ \bibnamefont {Averitt}}, \bibinfo {author} {\bibfnamefont {D.}~\bibnamefont {van~der Marel}}, \bibinfo {author} {\bibfnamefont {M.}~\bibnamefont {Dressel}},\ and\ \bibinfo {author} {\bibfnamefont {K.}~\bibnamefont {Haule}},\ }\bibfield  {title} {\bibinfo {title} {{Electrodynamics of correlated electron materials}},\ }\href {https://doi.org/10.1103/RevModPhys.83.471} {\bibfield  {journal} {\bibinfo  {journal} {Rev. Mod. Phys.}\ }\textbf {\bibinfo {volume} {83}},\ \bibinfo {pages} {471} (\bibinfo {year} {2011})}\BibitemShut {NoStop}%
\bibitem [{\citenamefont {Basov}\ and\ \citenamefont {Timusk}(2005)}]{Basov2005}%
  \BibitemOpen
  \bibfield  {author} {\bibinfo {author} {\bibfnamefont {D.~N.}\ \bibnamefont {Basov}}\ and\ \bibinfo {author} {\bibfnamefont {T.}~\bibnamefont {Timusk}},\ }\bibfield  {title} {\bibinfo {title} {{Electrodynamics of high-${T}_{c}$ superconductors}},\ }\href {https://doi.org/10.1103/RevModPhys.77.721} {\bibfield  {journal} {\bibinfo  {journal} {Rev. Mod. Phys.}\ }\textbf {\bibinfo {volume} {77}},\ \bibinfo {pages} {721} (\bibinfo {year} {2005})}\BibitemShut {NoStop}%
\bibitem [{\citenamefont {Burch}\ \emph {et~al.}(2018)\citenamefont {Burch}, \citenamefont {Mandrus},\ and\ \citenamefont {Park}}]{Burch2018}%
  \BibitemOpen
  \bibfield  {author} {\bibinfo {author} {\bibfnamefont {K.~S.}\ \bibnamefont {Burch}}, \bibinfo {author} {\bibfnamefont {D.}~\bibnamefont {Mandrus}},\ and\ \bibinfo {author} {\bibfnamefont {J.-G.}\ \bibnamefont {Park}},\ }\bibfield  {title} {\bibinfo {title} {{Magnetism in two-dimensional van der Waals materials}},\ }\href {https://doi.org/10.1038/s41586-018-0631-z} {\bibfield  {journal} {\bibinfo  {journal} {Nature}\ }\textbf {\bibinfo {volume} {563}},\ \bibinfo {pages} {47–52} (\bibinfo {year} {2018})}\BibitemShut {NoStop}%
\bibitem [{\citenamefont {Barbier}\ \emph {et~al.}(2022)\citenamefont {Barbier}, \citenamefont {Wilhelm}, \citenamefont {Colin}, \citenamefont {Opagiste}, \citenamefont {Lhotel}, \citenamefont {Pinek}, \citenamefont {Kim}, \citenamefont {Braithwaite}, \citenamefont {Ressouche}, \citenamefont {Ohresser}, \citenamefont {Otero}, \citenamefont {Rogalev},\ and\ \citenamefont {Ouisse}}]{Barbier2022}%
  \BibitemOpen
  \bibfield  {author} {\bibinfo {author} {\bibfnamefont {M.}~\bibnamefont {Barbier}}, \bibinfo {author} {\bibfnamefont {F.}~\bibnamefont {Wilhelm}}, \bibinfo {author} {\bibfnamefont {C.~V.}\ \bibnamefont {Colin}}, \bibinfo {author} {\bibfnamefont {C.}~\bibnamefont {Opagiste}}, \bibinfo {author} {\bibfnamefont {E.}~\bibnamefont {Lhotel}}, \bibinfo {author} {\bibfnamefont {D.}~\bibnamefont {Pinek}}, \bibinfo {author} {\bibfnamefont {Y.}~\bibnamefont {Kim}}, \bibinfo {author} {\bibfnamefont {D.}~\bibnamefont {Braithwaite}}, \bibinfo {author} {\bibfnamefont {E.}~\bibnamefont {Ressouche}}, \bibinfo {author} {\bibfnamefont {P.}~\bibnamefont {Ohresser}}, \bibinfo {author} {\bibfnamefont {E.}~\bibnamefont {Otero}}, \bibinfo {author} {\bibfnamefont {A.}~\bibnamefont {Rogalev}},\ and\ \bibinfo {author} {\bibfnamefont {T.}~\bibnamefont {Ouisse}},\ }\bibfield  {title} {\bibinfo {title} {{Magnetic properties of the $({\mathrm{Mo}}_{2/3}{R}_{1/3}){}_{2}\mathrm{AlC} (R=\mathrm{Ho},\mathrm{Dy})
  i\text{\ensuremath{-}}\mathrm{MAX}$ phases studied by x-ray magnetic circular dichroism and neutron diffraction}},\ }\href {https://doi.org/10.1103/PhysRevB.105.174421} {\bibfield  {journal} {\bibinfo  {journal} {Phys. Rev. B}\ }\textbf {\bibinfo {volume} {105}},\ \bibinfo {pages} {174421} (\bibinfo {year} {2022})}\BibitemShut {NoStop}%
\bibitem [{\citenamefont {Champagne}\ \emph {et~al.}(2019)\citenamefont {Champagne}, \citenamefont {Chaix-Pluchery}, \citenamefont {Ouisse}, \citenamefont {Pinek}, \citenamefont {G\'elard}, \citenamefont {Jouffret}, \citenamefont {Barbier}, \citenamefont {Wilhelm}, \citenamefont {Tao}, \citenamefont {Lu}, \citenamefont {Rosen}, \citenamefont {Barsoum},\ and\ \citenamefont {Charlier}}]{Champagne2019}%
  \BibitemOpen
  \bibfield  {author} {\bibinfo {author} {\bibfnamefont {A.}~\bibnamefont {Champagne}}, \bibinfo {author} {\bibfnamefont {O.}~\bibnamefont {Chaix-Pluchery}}, \bibinfo {author} {\bibfnamefont {T.}~\bibnamefont {Ouisse}}, \bibinfo {author} {\bibfnamefont {D.}~\bibnamefont {Pinek}}, \bibinfo {author} {\bibfnamefont {I.}~\bibnamefont {G\'elard}}, \bibinfo {author} {\bibfnamefont {L.}~\bibnamefont {Jouffret}}, \bibinfo {author} {\bibfnamefont {M.}~\bibnamefont {Barbier}}, \bibinfo {author} {\bibfnamefont {F.}~\bibnamefont {Wilhelm}}, \bibinfo {author} {\bibfnamefont {Q.}~\bibnamefont {Tao}}, \bibinfo {author} {\bibfnamefont {J.}~\bibnamefont {Lu}}, \bibinfo {author} {\bibfnamefont {J.}~\bibnamefont {Rosen}}, \bibinfo {author} {\bibfnamefont {M.~W.}\ \bibnamefont {Barsoum}},\ and\ \bibinfo {author} {\bibfnamefont {J.-C.}\ \bibnamefont {Charlier}},\ }\bibfield  {title} {\bibinfo {title} {{First-order Raman scattering of rare-earth containing $i$-MAX single crystals
  ${({\mathrm{Mo}}_{2/3}{\mathrm{RE}}_{1/3})}_{2}\mathrm{AlC} (\mathrm{RE}=\mathrm{Nd},\phantom{\rule{0.16em}{0ex}}\mathrm{Gd},\phantom{\rule{0.16em}{0ex}}\mathrm{Dy},\phantom{\rule{0.16em}{0ex}}\mathrm{Ho},\phantom{\rule{0.16em}{0ex}}\mathrm{Er})$}},\ }\href {https://doi.org/10.1103/PhysRevMaterials.3.053609} {\bibfield  {journal} {\bibinfo  {journal} {Phys. Rev. Mater.}\ }\textbf {\bibinfo {volume} {3}},\ \bibinfo {pages} {053609} (\bibinfo {year} {2019})}\BibitemShut {NoStop}%
\bibitem [{\citenamefont {Kuzmenko}(2005)}]{Kuzmenko2005}%
  \BibitemOpen
  \bibfield  {author} {\bibinfo {author} {\bibfnamefont {A.~B.}\ \bibnamefont {Kuzmenko}},\ }\bibfield  {title} {\bibinfo {title} {{Kramers–Kronig constrained variational analysis of optical spectra}},\ }\href {https://doi.org/10.1063/1.1979470} {\bibfield  {journal} {\bibinfo  {journal} {Review of Scientific Instruments}\ }\textbf {\bibinfo {volume} {76}},\ \bibinfo {pages} {1979470} (\bibinfo {year} {2005})}\BibitemShut {NoStop}%
\bibitem [{\citenamefont {Cashion}\ \emph {et~al.}(1968)\citenamefont {Cashion}, \citenamefont {Cooke}, \citenamefont {Hawkes}, \citenamefont {Leask}, \citenamefont {Thorp},\ and\ \citenamefont {Wells}}]{Cashion1968}%
  \BibitemOpen
  \bibfield  {author} {\bibinfo {author} {\bibfnamefont {J.~D.}\ \bibnamefont {Cashion}}, \bibinfo {author} {\bibfnamefont {A.~H.}\ \bibnamefont {Cooke}}, \bibinfo {author} {\bibfnamefont {J.~F.~B.}\ \bibnamefont {Hawkes}}, \bibinfo {author} {\bibfnamefont {M.~J.~M.}\ \bibnamefont {Leask}}, \bibinfo {author} {\bibfnamefont {T.~L.}\ \bibnamefont {Thorp}},\ and\ \bibinfo {author} {\bibfnamefont {M.~R.}\ \bibnamefont {Wells}},\ }\bibfield  {title} {\bibinfo {title} {{Magnetic Properties of Antiferromagnetic GdAlO$_{3}$}},\ }\href {https://doi.org/10.1063/1.1656304} {\bibfield  {journal} {\bibinfo  {journal} {Journal of Applied Physics}\ }\textbf {\bibinfo {volume} {39}},\ \bibinfo {pages} {1360–1361} (\bibinfo {year} {1968})}\BibitemShut {NoStop}%
\bibitem [{\citenamefont {Tao}\ \emph {et~al.}(2019)\citenamefont {Tao}, \citenamefont {Lu}, \citenamefont {Dahlqvist}, \citenamefont {Mockute}, \citenamefont {Calder}, \citenamefont {Petruhins}, \citenamefont {Meshkian}, \citenamefont {Rivin}, \citenamefont {Potashnikov}, \citenamefont {Caspi}, \citenamefont {Shaked}, \citenamefont {Hoser}, \citenamefont {Opagiste}, \citenamefont {Galera}, \citenamefont {Salikhov}, \citenamefont {Wiedwald}, \citenamefont {Ritter}, \citenamefont {Wildes}, \citenamefont {Johansson}, \citenamefont {Hultman}, \citenamefont {Farle}, \citenamefont {Barsoum},\ and\ \citenamefont {Rosen}}]{Tao2019}%
  \BibitemOpen
  \bibfield  {author} {\bibinfo {author} {\bibfnamefont {Q.}~\bibnamefont {Tao}}, \bibinfo {author} {\bibfnamefont {J.}~\bibnamefont {Lu}}, \bibinfo {author} {\bibfnamefont {M.}~\bibnamefont {Dahlqvist}}, \bibinfo {author} {\bibfnamefont {A.}~\bibnamefont {Mockute}}, \bibinfo {author} {\bibfnamefont {S.}~\bibnamefont {Calder}}, \bibinfo {author} {\bibfnamefont {A.}~\bibnamefont {Petruhins}}, \bibinfo {author} {\bibfnamefont {R.}~\bibnamefont {Meshkian}}, \bibinfo {author} {\bibfnamefont {O.}~\bibnamefont {Rivin}}, \bibinfo {author} {\bibfnamefont {D.}~\bibnamefont {Potashnikov}}, \bibinfo {author} {\bibfnamefont {E.~N.}\ \bibnamefont {Caspi}}, \bibinfo {author} {\bibfnamefont {H.}~\bibnamefont {Shaked}}, \bibinfo {author} {\bibfnamefont {A.}~\bibnamefont {Hoser}}, \bibinfo {author} {\bibfnamefont {C.}~\bibnamefont {Opagiste}}, \bibinfo {author} {\bibfnamefont {R.-M.}\ \bibnamefont {Galera}}, \bibinfo {author} {\bibfnamefont {R.}~\bibnamefont {Salikhov}}, \bibinfo {author} {\bibfnamefont {U.}~\bibnamefont
  {Wiedwald}}, \bibinfo {author} {\bibfnamefont {C.}~\bibnamefont {Ritter}}, \bibinfo {author} {\bibfnamefont {A.~R.}\ \bibnamefont {Wildes}}, \bibinfo {author} {\bibfnamefont {B.}~\bibnamefont {Johansson}}, \bibinfo {author} {\bibfnamefont {L.}~\bibnamefont {Hultman}}, \bibinfo {author} {\bibfnamefont {M.}~\bibnamefont {Farle}}, \bibinfo {author} {\bibfnamefont {M.~W.}\ \bibnamefont {Barsoum}},\ and\ \bibinfo {author} {\bibfnamefont {J.}~\bibnamefont {Rosen}},\ }\bibfield  {title} {\bibinfo {title} {{Atomically Layered and Ordered Rare-Earth i-MAX Phases: A New Class of Magnetic Quaternary Compounds}},\ }\href {https://doi.org/10.1021/acs.chemmater.8b05298} {\bibfield  {journal} {\bibinfo  {journal} {Chemistry of Materials}\ }\textbf {\bibinfo {volume} {31}},\ \bibinfo {pages} {2476–2485} (\bibinfo {year} {2019})}\BibitemShut {NoStop}%
\bibitem [{\citenamefont {Levin}\ \emph {et~al.}(2001)\citenamefont {Levin}, \citenamefont {Pecharsky},\ and\ \citenamefont {Gschneidner}}]{Levin2001}%
  \BibitemOpen
  \bibfield  {author} {\bibinfo {author} {\bibfnamefont {E.~M.}\ \bibnamefont {Levin}}, \bibinfo {author} {\bibfnamefont {V.~K.}\ \bibnamefont {Pecharsky}},\ and\ \bibinfo {author} {\bibfnamefont {K.~A.}\ \bibnamefont {Gschneidner}},\ }\bibfield  {title} {\bibinfo {title} {{Real and imaginary components of the alternating current magnetic susceptibility of RAl$_{2}$ (R=Gd, Dy, and Er) in the ferromagnetic region}},\ }\href {https://doi.org/10.1063/1.1415056} {\bibfield  {journal} {\bibinfo  {journal} {Journal of Applied Physics}\ }\textbf {\bibinfo {volume} {90}},\ \bibinfo {pages} {6255–6262} (\bibinfo {year} {2001})}\BibitemShut {NoStop}%
\bibitem [{\citenamefont {Hashemi}\ \emph {et~al.}(2017)\citenamefont {Hashemi}, \citenamefont {Komsa}, \citenamefont {Puska},\ and\ \citenamefont {Krasheninnikov}}]{Hashemi2017}%
  \BibitemOpen
  \bibfield  {author} {\bibinfo {author} {\bibfnamefont {A.}~\bibnamefont {Hashemi}}, \bibinfo {author} {\bibfnamefont {H.-P.}\ \bibnamefont {Komsa}}, \bibinfo {author} {\bibfnamefont {M.}~\bibnamefont {Puska}},\ and\ \bibinfo {author} {\bibfnamefont {A.~V.}\ \bibnamefont {Krasheninnikov}},\ }\bibfield  {title} {\bibinfo {title} {{Vibrational Properties of Metal Phosphorus Trichalcogenides from First-Principles Calculations}},\ }\href {https://doi.org/10.1021/acs.jpcc.7b09634} {\bibfield  {journal} {\bibinfo  {journal} {The Journal of Physical Chemistry C}\ }\textbf {\bibinfo {volume} {121}},\ \bibinfo {pages} {27207–27217} (\bibinfo {year} {2017})}\BibitemShut {NoStop}%
\bibitem [{\citenamefont {Wyzula}\ \emph {et~al.}(2022)\citenamefont {Wyzula}, \citenamefont {Mohelský}, \citenamefont {Václavková}, \citenamefont {Kapuscinski}, \citenamefont {Veis}, \citenamefont {Faugeras}, \citenamefont {Potemski}, \citenamefont {Zhitomirsky},\ and\ \citenamefont {Orlita}}]{Wyzula2022}%
  \BibitemOpen
  \bibfield  {author} {\bibinfo {author} {\bibfnamefont {J.}~\bibnamefont {Wyzula}}, \bibinfo {author} {\bibfnamefont {I.}~\bibnamefont {Mohelský}}, \bibinfo {author} {\bibfnamefont {D.}~\bibnamefont {Václavková}}, \bibinfo {author} {\bibfnamefont {P.}~\bibnamefont {Kapuscinski}}, \bibinfo {author} {\bibfnamefont {M.}~\bibnamefont {Veis}}, \bibinfo {author} {\bibfnamefont {C.}~\bibnamefont {Faugeras}}, \bibinfo {author} {\bibfnamefont {M.}~\bibnamefont {Potemski}}, \bibinfo {author} {\bibfnamefont {M.~E.}\ \bibnamefont {Zhitomirsky}},\ and\ \bibinfo {author} {\bibfnamefont {M.}~\bibnamefont {Orlita}},\ }\bibfield  {title} {\bibinfo {title} {{High-Angular Momentum Excitations in Collinear Antiferromagnet FePS$_{3}$}},\ }\href {https://doi.org/10.1021/acs.nanolett.2c04111} {\bibfield  {journal} {\bibinfo  {journal} {Nano Letters}\ }\textbf {\bibinfo {volume} {22}},\ \bibinfo {pages} {9741–9747} (\bibinfo {year} {2022})}\BibitemShut {NoStop}%
\bibitem [{\citenamefont {Gong}\ \emph {et~al.}(2017)\citenamefont {Gong}, \citenamefont {Li}, \citenamefont {Li}, \citenamefont {Ji}, \citenamefont {Stern}, \citenamefont {Xia}, \citenamefont {Cao}, \citenamefont {Bao}, \citenamefont {Wang}, \citenamefont {Wang}, \citenamefont {Qiu}, \citenamefont {Cava}, \citenamefont {Louie}, \citenamefont {Xia},\ and\ \citenamefont {Zhang}}]{Gong2017}%
  \BibitemOpen
  \bibfield  {author} {\bibinfo {author} {\bibfnamefont {C.}~\bibnamefont {Gong}}, \bibinfo {author} {\bibfnamefont {L.}~\bibnamefont {Li}}, \bibinfo {author} {\bibfnamefont {Z.}~\bibnamefont {Li}}, \bibinfo {author} {\bibfnamefont {H.}~\bibnamefont {Ji}}, \bibinfo {author} {\bibfnamefont {A.}~\bibnamefont {Stern}}, \bibinfo {author} {\bibfnamefont {Y.}~\bibnamefont {Xia}}, \bibinfo {author} {\bibfnamefont {T.}~\bibnamefont {Cao}}, \bibinfo {author} {\bibfnamefont {W.}~\bibnamefont {Bao}}, \bibinfo {author} {\bibfnamefont {C.}~\bibnamefont {Wang}}, \bibinfo {author} {\bibfnamefont {Y.}~\bibnamefont {Wang}}, \bibinfo {author} {\bibfnamefont {Z.~Q.}\ \bibnamefont {Qiu}}, \bibinfo {author} {\bibfnamefont {R.~J.}\ \bibnamefont {Cava}}, \bibinfo {author} {\bibfnamefont {S.~G.}\ \bibnamefont {Louie}}, \bibinfo {author} {\bibfnamefont {J.}~\bibnamefont {Xia}},\ and\ \bibinfo {author} {\bibfnamefont {X.}~\bibnamefont {Zhang}},\ }\bibfield  {title} {\bibinfo {title} {{Discovery of intrinsic ferromagnetism in
  two-dimensional van der Waals crystals}},\ }\href {https://doi.org/10.1038/nature22060} {\bibfield  {journal} {\bibinfo  {journal} {Nature}\ }\textbf {\bibinfo {volume} {546}},\ \bibinfo {pages} {265–269} (\bibinfo {year} {2017})}\BibitemShut {NoStop}%
\bibitem [{\citenamefont {Potashnikov}\ \emph {et~al.}(2021)\citenamefont {Potashnikov}, \citenamefont {Caspi}, \citenamefont {Pesach}, \citenamefont {Tao}, \citenamefont {Rosen}, \citenamefont {Sheptyakov}, \citenamefont {Evans}, \citenamefont {Ritter}, \citenamefont {Salman}, \citenamefont {Bonfa}, \citenamefont {Ouisse}, \citenamefont {Barbier}, \citenamefont {Rivin},\ and\ \citenamefont {Keren}}]{Potashnikov2021}%
  \BibitemOpen
  \bibfield  {author} {\bibinfo {author} {\bibfnamefont {D.}~\bibnamefont {Potashnikov}}, \bibinfo {author} {\bibfnamefont {E.~N.}\ \bibnamefont {Caspi}}, \bibinfo {author} {\bibfnamefont {A.}~\bibnamefont {Pesach}}, \bibinfo {author} {\bibfnamefont {Q.}~\bibnamefont {Tao}}, \bibinfo {author} {\bibfnamefont {J.}~\bibnamefont {Rosen}}, \bibinfo {author} {\bibfnamefont {D.}~\bibnamefont {Sheptyakov}}, \bibinfo {author} {\bibfnamefont {H.~A.}\ \bibnamefont {Evans}}, \bibinfo {author} {\bibfnamefont {C.}~\bibnamefont {Ritter}}, \bibinfo {author} {\bibfnamefont {Z.}~\bibnamefont {Salman}}, \bibinfo {author} {\bibfnamefont {P.}~\bibnamefont {Bonfa}}, \bibinfo {author} {\bibfnamefont {T.}~\bibnamefont {Ouisse}}, \bibinfo {author} {\bibfnamefont {M.}~\bibnamefont {Barbier}}, \bibinfo {author} {\bibfnamefont {O.}~\bibnamefont {Rivin}},\ and\ \bibinfo {author} {\bibfnamefont {A.}~\bibnamefont {Keren}},\ }\bibfield  {title} {\bibinfo {title} {{Magnetic structure determination of high-moment rare-earth-based
  laminates}},\ }\href {https://doi.org/10.1103/physrevb.104.174440} {\bibfield  {journal} {\bibinfo  {journal} {Physical Review B}\ }\textbf {\bibinfo {volume} {104}},\ \bibinfo {pages} {174440} (\bibinfo {year} {2021})}\BibitemShut {NoStop}%
\bibitem [{\citenamefont {Chen}\ \emph {et~al.}(1995)\citenamefont {Chen}, \citenamefont {Irwin},\ and\ \citenamefont {Franck}}]{Chen1995}%
  \BibitemOpen
  \bibfield  {author} {\bibinfo {author} {\bibfnamefont {X.~K.}\ \bibnamefont {Chen}}, \bibinfo {author} {\bibfnamefont {J.~C.}\ \bibnamefont {Irwin}},\ and\ \bibinfo {author} {\bibfnamefont {J.~P.}\ \bibnamefont {Franck}},\ }\bibfield  {title} {\bibinfo {title} {{Evidence for a strong spin-phonon interaction in cupric oxide}},\ }\href {https://doi.org/10.1103/physrevb.52.r13130} {\bibfield  {journal} {\bibinfo  {journal} {Physical Review B}\ }\textbf {\bibinfo {volume} {52}},\ \bibinfo {pages} {R13130–R13133} (\bibinfo {year} {1995})}\BibitemShut {NoStop}%
\bibitem [{\citenamefont {Chittari}\ \emph {et~al.}(2016)\citenamefont {Chittari}, \citenamefont {Park}, \citenamefont {Lee}, \citenamefont {Han}, \citenamefont {MacDonald}, \citenamefont {Hwang},\ and\ \citenamefont {Jung}}]{Chittari2016}%
  \BibitemOpen
  \bibfield  {author} {\bibinfo {author} {\bibfnamefont {B.~L.}\ \bibnamefont {Chittari}}, \bibinfo {author} {\bibfnamefont {Y.}~\bibnamefont {Park}}, \bibinfo {author} {\bibfnamefont {D.}~\bibnamefont {Lee}}, \bibinfo {author} {\bibfnamefont {M.}~\bibnamefont {Han}}, \bibinfo {author} {\bibfnamefont {A.~H.}\ \bibnamefont {MacDonald}}, \bibinfo {author} {\bibfnamefont {E.}~\bibnamefont {Hwang}},\ and\ \bibinfo {author} {\bibfnamefont {J.}~\bibnamefont {Jung}},\ }\bibfield  {title} {\bibinfo {title} {{Electronic and magnetic properties of single-layer $M\mathrm{P}{X}_{3}$ metal phosphorous trichalcogenides}},\ }\href {https://doi.org/10.1103/PhysRevB.94.184428} {\bibfield  {journal} {\bibinfo  {journal} {Phys. Rev. B}\ }\textbf {\bibinfo {volume} {94}},\ \bibinfo {pages} {184428} (\bibinfo {year} {2016})}\BibitemShut {NoStop}%
\bibitem [{\citenamefont {Coak}\ \emph {et~al.}(2021)\citenamefont {Coak}, \citenamefont {Jarvis}, \citenamefont {Hamidov}, \citenamefont {Wildes}, \citenamefont {Paddison}, \citenamefont {Liu}, \citenamefont {Haines}, \citenamefont {Dang}, \citenamefont {Kichanov}, \citenamefont {Savenko}, \citenamefont {Lee}, \citenamefont {Kratochv\'{\i}lov\'a}, \citenamefont {Klotz}, \citenamefont {Hansen}, \citenamefont {Kozlenko}, \citenamefont {Park},\ and\ \citenamefont {Saxena}}]{Coak2021}%
  \BibitemOpen
  \bibfield  {author} {\bibinfo {author} {\bibfnamefont {M.~J.}\ \bibnamefont {Coak}}, \bibinfo {author} {\bibfnamefont {D.~M.}\ \bibnamefont {Jarvis}}, \bibinfo {author} {\bibfnamefont {H.}~\bibnamefont {Hamidov}}, \bibinfo {author} {\bibfnamefont {A.~R.}\ \bibnamefont {Wildes}}, \bibinfo {author} {\bibfnamefont {J.~A.~M.}\ \bibnamefont {Paddison}}, \bibinfo {author} {\bibfnamefont {C.}~\bibnamefont {Liu}}, \bibinfo {author} {\bibfnamefont {C.~R.~S.}\ \bibnamefont {Haines}}, \bibinfo {author} {\bibfnamefont {N.~T.}\ \bibnamefont {Dang}}, \bibinfo {author} {\bibfnamefont {S.~E.}\ \bibnamefont {Kichanov}}, \bibinfo {author} {\bibfnamefont {B.~N.}\ \bibnamefont {Savenko}}, \bibinfo {author} {\bibfnamefont {S.}~\bibnamefont {Lee}}, \bibinfo {author} {\bibfnamefont {M.}~\bibnamefont {Kratochv\'{\i}lov\'a}}, \bibinfo {author} {\bibfnamefont {S.}~\bibnamefont {Klotz}}, \bibinfo {author} {\bibfnamefont {T.~C.}\ \bibnamefont {Hansen}}, \bibinfo {author} {\bibfnamefont {D.~P.}\ \bibnamefont {Kozlenko}}, \bibinfo
  {author} {\bibfnamefont {J.-G.}\ \bibnamefont {Park}},\ and\ \bibinfo {author} {\bibfnamefont {S.~S.}\ \bibnamefont {Saxena}},\ }\bibfield  {title} {\bibinfo {title} {{Emergent Magnetic Phases in Pressure-Tuned van der Waals Antiferromagnet ${\mathrm{FePS}}_{3}$}},\ }\href {https://doi.org/10.1103/PhysRevX.11.011024} {\bibfield  {journal} {\bibinfo  {journal} {Phys. Rev. X}\ }\textbf {\bibinfo {volume} {11}},\ \bibinfo {pages} {011024} (\bibinfo {year} {2021})}\BibitemShut {NoStop}%
\bibitem [{\citenamefont {Pawbake}\ \emph {et~al.}(2022)\citenamefont {Pawbake}, \citenamefont {Pelini}, \citenamefont {Delhomme}, \citenamefont {Romanin}, \citenamefont {Vaclavkova}, \citenamefont {Martinez}, \citenamefont {Calandra}, \citenamefont {Measson}, \citenamefont {Veis}, \citenamefont {Potemski}, \citenamefont {Orlita},\ and\ \citenamefont {Faugeras}}]{Pawbake2022}%
  \BibitemOpen
  \bibfield  {author} {\bibinfo {author} {\bibfnamefont {A.}~\bibnamefont {Pawbake}}, \bibinfo {author} {\bibfnamefont {T.}~\bibnamefont {Pelini}}, \bibinfo {author} {\bibfnamefont {A.}~\bibnamefont {Delhomme}}, \bibinfo {author} {\bibfnamefont {D.}~\bibnamefont {Romanin}}, \bibinfo {author} {\bibfnamefont {D.}~\bibnamefont {Vaclavkova}}, \bibinfo {author} {\bibfnamefont {G.}~\bibnamefont {Martinez}}, \bibinfo {author} {\bibfnamefont {M.}~\bibnamefont {Calandra}}, \bibinfo {author} {\bibfnamefont {M.-A.}\ \bibnamefont {Measson}}, \bibinfo {author} {\bibfnamefont {M.}~\bibnamefont {Veis}}, \bibinfo {author} {\bibfnamefont {M.}~\bibnamefont {Potemski}}, \bibinfo {author} {\bibfnamefont {M.}~\bibnamefont {Orlita}},\ and\ \bibinfo {author} {\bibfnamefont {C.}~\bibnamefont {Faugeras}},\ }\bibfield  {title} {\bibinfo {title} {{High-Pressure Tuning of Magnon-Polarons in the Layered Antiferromagnet FePS$_{3}$}},\ }\href {https://doi.org/10.1021/acsnano.2c04286} {\bibfield  {journal} {\bibinfo  {journal} {ACS Nano}\
  }\textbf {\bibinfo {volume} {16}},\ \bibinfo {pages} {12656–12665} (\bibinfo {year} {2022})}\BibitemShut {NoStop}%
\end{thebibliography}%

\end{document}